\documentclass[12pt]{article}
\usepackage{a4wide}
\usepackage[dvips]{graphicx,psfrag}
\usepackage{amsfonts}
\usepackage{amstext}
\usepackage{amsmath}
\usepackage{amssymb}
\usepackage{amsthm}
\usepackage[mathscr]{eucal}
\makeatletter
 
 \@addtoreset{equation}{section}
\makeatother
\def\a{\alpha}
\def\A{\mathcal{A}}
\def\Ai{{\rm Ai}}
\def\b{\beta}
\def\B{\mathcal{B}}

\def\D{\mathcal{D}}

\def\e{\epsilon}

\def\L{\mathcal{L}}

\def\o{\omega}

\def\P{\mathbb{P}}
\def\1{\bf{1}}
\def\R{\mathbb{R}}

\begin{document}
\title{Stationary correlations for the 1D KPZ equation}
\author{Takashi Imamura
\footnote { Research Center for Advanced Science and Technology
, The University of Tokyo,~E-mail:imamura@jamology.rcast.u-tokyo.ac.jp}
, Tomohiro Sasamoto
\footnote { Department of mathematics and informatics, 
Chiba University,~E-mail: sasamoto@math.s.chiba-u.ac.jp}}

\maketitle

\begin{abstract}
We study exact stationary properties of the one-dimensional 
Kardar-Parisi-Zhang (KPZ) equation by using the replica approach. 
The stationary state for the KPZ equation is realized by setting the
initial condition the two-sided Brownian motion (BM) with respect to
the space variable.
Developing techniques for dealing with this initial condition 
in the replica analysis, we elucidate some exact nature of 
the height fluctuation for the KPZ equation. In particular, we obtain an explicit representation of the probability distribution of the height in terms of the Fredholm determinants. Furthermore from this expression, we also get 
the exact expression of the space-time two-point correlation function. 
\end{abstract}

\section{Introduction}
Surface growth phenomena have attracted much attention in both
science and technology, which involve fundamental understanding of 
the roughness of the surface and applications to the design of
new materials. Being typical nonlinear and nonequilibrium phenomena,
they have been one of the main topics in nonequilibrium statistical
physics. In 1986, Kardar, Parisi and Zhang proposed a prototypical 
equation which describes surface growth with local interaction~\cite{KPZ1986}. 
Its one-dimensional version reads
\begin{equation}
\frac{\partial h(x,t)}{\partial t}=\frac{\lambda}{2}
\left(\frac{\partial h(x,t)}{\partial x}\right)^2+
\nu\frac{\partial^2 h(x,t)}{\partial x^2}+\sqrt{D}\eta(x,t).
\label{KPZ}
\end{equation}
Here $h(x,t)$ represents the height of the surface at position $x\in \R$ and 
time $t\ge 0$. The first term represents the effect of nonlinearity and 
the second one describes the smoothing effect of the surface. $\eta (x,t)$ 
represents randomness described by the Gaussian white noise with covariance,
\begin{equation}
\langle \eta(x,t)\eta(x',t')\rangle=\delta(x-x')\delta(t-t').
\label{random}
\end{equation}
The parameters $\lambda,~\nu,~D$ determine their strengths. 

An important feature of the KPZ equation is that it can describe 
self-affine properties of the height fluctuation. In one-dimensional case, 
it has been shown by the renormalization techniques that the height 
fluctuation grows like $t^{1/3}$ as $t$ goes to infinity~\cite{KPZ1986}. 
This {\it growth exponent} $1/3$ appear also in many models of surface growth and characterizes the KPZ universality class~\cite{BS1995,Meakin1998}. 

Our understanding of the KPZ universality class has progressed 
by finding mathematical connection between the probability distribution 
of the height and the random matrix theory.
The first breakthrough was brought not in the KPZ equation
but in solvable discrete models belonging to the KPZ universality class 
such as the totally asymmetric simple exclusion process (TASEP) 
and the polynuclear 
growth (PNG) model. In~\cite{Jo2000}, the author obtained  the exact solution 
of the current distribution for the TASEP with the step initial configuration, 
which corresponds to the height distribution for the curved growth 
in the language of growth process. It turned out that 
it converges in the long time limit to the largest eigenvalue distribution 
of the Gaussian Unitary Ensemble (GUE) in the random matrix theory. This distribution is called 
the GUE Tracy-Widom distribution~\cite{TW1994}. 
The result in~\cite{Jo2000} has been generalized 
to the case of the other initial 
configurations~\cite{PS2000b,BR2000} and it has been found that
the current distributions depend on the initial conditions even if
they have the common scaling exponents $1/3$.
For example, in the case of the 
alternating initial condition for TASEP, 
which corresponds to the flat initial 
configuration in a growth process, 
the limiting current distribution converges 
to the Gaussian Orthogonal Ensemble (GOE) Tracy-Widom distribution, 
which is the largest eigenvalue distribution of the other 
random matrix ensemble called the GOE~\cite{PS2000b,Sa2005,BFPS2007}. 
Since then, the studies on the height distributions have been 
the subject of active investigation~\cite{S2007r,F2010r,KK2010r}.

During the last few years, the studies on the KPZ equation and 
KPZ universality have entered a new stage~\cite{SS2010r,C2011r}. First,
an experiment using the liquid crystal turbulence were 
performed in~\cite{TS2010,TSSS2011,TS2012}. The authors succeeded in obtaining 
not only the exponent $1/3$ but also the GUE/GOE Tracy-Widom distributions 
as well as their finite-time corrections.  Second, there has been 
much progress on the theoretical study:
An exact height distribution for the KPZ equation
has been found in \cite{SS2010a,SS2010b,SS2010c,ACQ2010}.
This first exact solution was obtained for the narrow wedge initial condition, 
from which the surface grows in parabolic shape. 
The analysis was based on the result of the current distribution
of the ASEP in~\cite{TW2008,TW2009} and the fact that the KPZ equation 
is obtained as a weakly asymmetric limit of the ASEP~\cite{BG1997}. 
Afterwards the same technique was applied to the case of the half 
Brownian motion initial condition~\cite{CQ2010p}.
A remarkable feature of this solution is that this distribution
describes the universal crossover between 
the Edwards-Wilkinson~\cite{EW1982} and 
the KPZ universality classes. As $t\rightarrow 0$, the growth exponent becomes
$1/4$ and the distribution converges to the Gaussian, which implies that the
nonlinearity in the KPZ equation does not become effective and thus 
the system belongs to the Edwards-Wilkinson universality class.
As $t\rightarrow \infty$, on the other hand, the exponent becomes $1/3$ and 
the distribution converges to the GUE Tracy-Widom distribution. 
Furthermore it turned out that this crossover is universal for weakly 
driven growth~\cite{AKQ2010,CLDR2010}.

As explained above, these achievements in the KPZ equation provide us 
not only exact information but also the important physical findings.
Thus it is of great importance to develop the exact approach in such a way that
it can be applied to other interesting situations. In particular, 
a most challenging problem would be an application of the exact solution 
to the analysis on the space-time correlation 
in the case of the stationary situation, which is 
the main subject of this paper. For this purpose, the replica method 
is suitable.
The replica method has been established
as a powerful tool for disordered systems in statistical mechanics. 
The application to the KPZ equation was first proposed 
in~\cite{K1987} and refined in~\cite{CLDR2010,D2010,D2010p2}.
It utilizes a connection between the $N$th moment of the exponential height
($N$ replica partition function) and the dynamics of one-dimensional $N$ particle Bose gas system with attractive interaction which can be solved
by the Bethe ansatz~\cite{LL1963,M1964}.
Using the exact knowledge to the $N$ replica partition function obtained by
the Bethe ansatz technique, we analyze the generating function of 
the $N$ replica partition function.
Although it is in fact a divergent sum, 
a resummation technique developed in~\cite{CLDR2010,D2010,D2010p2} allows us to
find Fredholm determinant representation of the generating function from which
the exact height distribution is readily obtained.
The method is so powerful that not merely the same expression 
was rederived for the narrow wedge initial condition~\cite{CLDR2010,
D2010,D2010p2},
but one can generalize it to the case of other initial conditions, 
such as the flat~\cite{CLD2011,CLD2012p} and half-Brownian motion~\cite{IS2011}
initial condition. 
The multi-point distributions have
also been studied~\cite{PS2010p,PS2010p2,IS2011} though it includes 
a decoupling assumption. Furthermore the idea of finding 
the Fredholm determinant structure of moment generating function has 
become the basis of the study on the Macdonald 
process~\cite{BC2011p}, which is a larger class of 
integrable stochastic process 
including the KPZ equation, a directed polymer model~\cite{O2012}, etc. 
and deals with the rigorous version of the
replica analysis. For more recent developments, see~\cite{BCF2012p,BCS2012p}.

In this paper, we consider the stationary situation of the KPZ equation by 
the replica method.  
It is known that the stationary state 
is given by the Brownian motion (BM)~\cite{KS1992} with respect to
the position ($x$) axis i.e., the stationary state is realized when we prepare
the initial height difference $h(x,0)-h(0,0)$ as the two-sided Brownian motion.
Since the system is translational invariant, one can set 
without loss of generality the initial height a the origin to be zero, i.e.,
$h(0,0)=0$. 
Then the initial condition considered in this paper is given by 
\begin{equation}
 \frac{\lambda}{2\nu}h(x,0) = \begin{cases}
	   \a B_-(-x), & x<0,\\
           \a B_+(x),  & x>0,
	  \end{cases}
\label{init_2bm}
\end{equation}
where $\alpha = (2\nu)^{-3/2} \lambda D^{1/2}$ and $B_{\pm}(x)$ are two independent standard BMs
with expectation $E[B(x)]=0$ and covariance 
$\text{Cov}[B(x)B(y)] = \text{min}\{x,y\}$. 
For this initial condition we consider the height distribution 
function
\begin{align}
\text{Prob}\left(h(x,t)\le s\right)
\label{heightf}
\end{align}
and obtain its explicit expression in terms of the Fredholm determinant. 
From this expression, we also obtain the stationary two point correlation 
function, 
\begin{align}
\langle ( h(x,t)-\langle h(x,t)\rangle)^2\rangle,
\label{2ptf}
\end{align}
which is one of the most fundamental quantities 
which detects the space-time correlation in statistical mechanical systems.
Note that~\eqref{2ptf} describes the correlation between $h(x,t)$ and $h(0,0)$
although $h(0,0)$ does not appear in this expression since we set $h(0,0)=0$
as explained above.
We also show that in the long-time limit, 
scaled forms of these functions~\eqref{heightf} 
and~\eqref{2ptf} converge to  the ones obtained in~\cite{BR2000} 
and~\cite{PS2002a,PS2004,FS2006,BFP2010} respectively.
A part of these result in this paper was already reported in~\cite{IS2012}.
The purpose of this paper is to provide complete descriptions of our results
including their derivations.

The paper is organized as follows. In Sec.~\ref{mmr}, we will give precise 
definitions of the quantities discussed in this paper and our results. 
In Sec.~\ref{ra}, the details of the replica analysis will be explained. 
In Sec.~\ref{gfhd}, we will obtain the exact solution for  
the generating function of the replica partition function and
the height distribution function for the two-sided Brownian motion 
initial condition with drifts. By taking the zero drift limit, 
our final goal, the height distribution for the stationary state, 
is accomplished, which will be explained in Sec.~\ref{sl}. The long-time
limit will be taken in  Sec.~\ref{lmt}. The last section will be devoted to
the conclusion.

\section{Model and main results}\label{mmr}
\subsection{Height distribution function}
\subsubsection{Two-sided Brownian motion initial condition with drifts}
\begin{figure}[h]
\begin{center}
\includegraphics[scale=0.6]{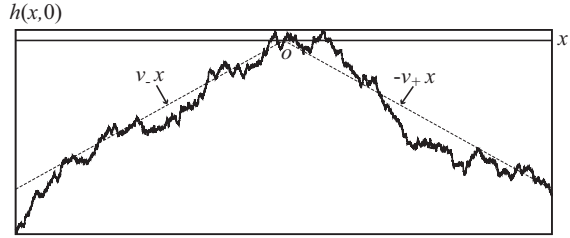}
\caption{\label{twosidedBM}
Two-sided Brownian motion initial condition with drifts $v_{\pm}$
}
\end{center}
\end{figure}

One of the goals in this paper is to obtain an exact expression 
for the distribution of the height~\eqref{heightf} 
with the two-sided Brownian motion initial condition~\eqref{init_2bm}. 
For this purpose, it is convenient as a first step to consider 
a slightly more generalized initial condition where we add the drift terms
to~\eqref{init_2bm},
\begin{align}
\frac{\lambda}{2\nu}h(x,0) =
\begin{cases}
\a B_-(-x) + \a^2v_- x, &x<0, \\
\a B_+(x) - \a^2v_+ x, &x>0, 
\end{cases}
\label{BMic}
\end{align}
where $v_{\pm}$ indicate the strengths of the drifts. 
Fig.~\ref{twosidedBM} depicts this generalized initial condition~\eqref{BMic}.
At first we set the drifts to be positive so that one can 
take an average over the Brownian motion initial condition
(see~\eqref{Betheexpand} below). Afterwards, the condition for
the drifts will be relaxed gradually as our analysis goes on
(see~\eqref{zndet} and~\eqref{aiggggex}). In Sec.~\ref{sl}, we finally
take the limit $v_{\pm}\rightarrow 0$ which corresponds to 
the stationary situation.

The macroscopic shape of the height $h(x,t)$ for large $t$
is obtained by solving~\eqref{KPZ} where only the nonlinear term
is taken into account and other terms (the diffusion and noise terms)
are ignored.  For the initial condition~\eqref{BMic}, it becomes
\begin{align}
\frac{\lambda}{2\nu}h(x,t)=
\begin{cases}
\a^2v_-x+\nu\a^4v_-^2t&~\text{for~} x\le -2\nu v_-\a^2t\\
-\frac{x^2}{4\nu t}&~\text{for~}\ -2\nu v_-\a^2t\le x\le
2\nu v_+\a^2t\\
-\a^2v_+x+\nu\a^4v_+^2t&~\text{for~} 2\nu v_+\a^2t\le x,
\end{cases}
\label{macroshape}
\end{align}
where $\a$ is given below \eqref{init_2bm}.
We are interested in the fluctuation property near the origin $x=0$ around
this macroscopic profile.
From the KPZ scaling, one expects that the fluctuation of 
the height scales like $O(t^{1/3})$ 
and nontrivial correlations are 
seen in the $x$ direction with scale $O(t^{2/3})$ when $t$ is large.  
Let us define a parameter $\gamma_t$ which scales as $O(t^{1/3})$ and 
a rescaled space coordinate $X$ by 
\begin{align}
\gamma_t=\left(\a^4 \nu t\right)^{\frac13},~~
x=\frac{2\gamma_t^2X}{\a^2}.
\label{generalgammaX}
\end{align}
We can introduce the scaled height $H_t(X)$ by 
\begin{equation}
\frac{\lambda}{2\nu}h 
\left( \frac{2\gamma_t^2X}{\a^2},t \right)
=
-\frac{\gamma_t^3}{12}-\gamma_tX^2 + \gamma_t H_t(X).
\label{scaled}
\end{equation}
Here the term $-\gamma_tX^2$ in RHS comes from 
the parabolic profile (the middle equation) in~\eqref{macroshape}.
Let us define the probability distribution of $H_t(X)$,
\begin{align}
{F}_{v_+,v_-,t}(s;X)=\text{Prob}\left(\frac{\lambda}{2\nu}{h}
\left(\frac{2\gamma_t^2X}{\a^2},t\right)+\frac{\gamma_t^3}{12} +\gamma_tX^2\le \gamma_t s \right)
=\text{Prob}({H}_t(X) \leq s).
\label{hdist}
\end{align}
In order to get an exact expression for this quantity,
we now introduce the generalized height $\tilde{h}(x,t)$ defined by 
\begin{align}
\tilde{h}(x,t)=h(x,t)+\chi,
\label{tildechi}
\end{align}
where the initial height of the overall surface $\chi$
is set to be a random variable such that $e^{\chi}$
obeys the inverse gamma distribution 
with parameter $v_+ + v_-$, i.e. the probability density function
of $e^\chi$ is given by
\begin{align}
p(x)=\frac{1}{\Gamma{(v_++v_-)}}x^{-(v_++v_-)-1}e^{-\frac{1}{x}}
{\bf{1}}_{x\ge 0}.
\label{invgamma}
\end{align}
Although this definition~\eqref{tildechi} may look artificial, 
we will find that it has more tractable mathematical structure
compared with $h(x,t)$ itself
(see Proposition 1 below and Sec.~\ref{Shifting}). 
We also define the probability distribution 
$\tilde{F}_{v_+,v_-,t}(s;X)$ in a similar way to~\eqref{hdist}
\begin{align}
\tilde{F}_{v_+,v_-,t}(s;X)=\text{Prob}\left(\frac{\lambda}{2\nu}
\tilde{h}\left(\frac{2\gamma_t^2X}{\a^2},t\right)+\frac{\gamma_t^3}{12} 
+\gamma_tX^2\le \gamma_t s \right)
=\text{Prob}({H}_t(X) +\frac{\lambda\chi}{2\nu\gamma_t}\leq s).
\label{thdist}
\end{align}

Our result for ${F}_{v_{+},v_-,t}(s;X)$ 
and $\tilde{F}_{v_{+},v_-,t}(s;X)$ are summarized 
as follows.

\vspace{5mm}
\noindent
{\bf Proposition 1}
{\it 
\begin{align}
&F_{v_+,v_-,t}(s)
=
\frac{\Gamma(v_++v_-)}{\Gamma(v_++v_-+\gamma_t^{-1}d/ds)}
\tilde{F}_{v_+,v_-,t}(s),
\label{Result}
\\
&\tilde{F}_{v_+,v_-,t}(s;X)
=
1-\int_{-\infty}^{\infty}
du e^{-e^{\gamma_t(s-u)}}
g_{\gamma_t}(u;X).
\label{Pr1}
\end{align}
Here $g_{\gamma_t}(u;X)$ is expressed as a difference of two
Fredholm determinants, 
\begin{align}
g_{\gamma_t}(u;X)
=\det\left(1-P_u(B^{\Gamma}_{\gamma_t}-P^{\Gamma}_{\Ai})P_u\right)
 -\det\left(1-P_uB^\Gamma_{\gamma_t}P_u\right),
\label{Th3g}
\end{align}
where $P_u$ is a projection operator on $(u,\infty)$, and
\begin{align}
&B^{\Gamma}_{\gamma_t}(\xi_1,\xi_2)
=\int_0^{\infty}dy
\Ai_\Gamma^\Gamma\left(\xi_1+y,
\frac{1}{\gamma_t},v_+-\frac{X}{\gamma_t},v_-+\frac{X}{\gamma_t}\right)
\Ai_\Gamma^\Gamma \left(\xi_2+y,
\frac{1}{\gamma_t},v_-+\frac{X}{\gamma_t},v_+-\frac{X}{\gamma_t}\right)
\notag\\
&+\int_{0}^{\infty}dy\frac{1}{e^{\gamma_t y}-1}
\left[
\Ai_\Gamma^\Gamma\left(\xi_1+y,
\frac{1}{\gamma_t},v_+-\frac{X}{\gamma_t},v_-+\frac{X}{\gamma_t}\right)
\Ai_\Gamma^\Gamma \left(\xi_2+y,
\frac{1}{\gamma_t},v_-+\frac{X}{\gamma_t},v_+-\frac{X}{\gamma_t}\right)
\right.\notag\\
&\hspace{2.8cm}\left.-
\Ai_\Gamma^\Gamma\left(\xi_1-y,
\frac{1}{\gamma_t},v_+-\frac{X}{\gamma_t},v_-+\frac{X}{\gamma_t}\right)
\Ai_\Gamma^\Gamma \left(\xi_2-y,
\frac{1}{\gamma_t},v_-+\frac{X}{\gamma_t},v_+-\frac{X}{\gamma_t}\right)
\right],
\label{BGamma}\\
&P^{\Gamma}_{\Ai}(\xi_1,\xi_2)=
\Ai_\Gamma^\Gamma\left(\xi_1,
\frac{1}{\gamma_t},v_+-\frac{X}{\gamma_t},v_-+\frac{X}{\gamma_t}\right)
\Ai_\Gamma^\Gamma \left(\xi_2,
\frac{1}{\gamma_t},v_-+\frac{X}{\gamma_t},v_+-\frac{X}{\gamma_t}\right),
\label{PGamma}
\end{align}
and the deformed Airy function $\Ai_{\Gamma}^{\Gamma}(a,b,c,d)$ is defined by
\begin{align}
&\Ai_{\Gamma}^\Gamma (a,b,c,d)
=\frac{1}{2\pi}
\int_{\Gamma_{i\frac{d}{b}}} dz 
e^{iza+i\frac{z^3}{3}}
\frac{\Gamma\left(ibz+d\right)}{\Gamma\left(-ibz+c\right)}.
\label{AiGGm}
\end{align}
Here
$\Gamma_{z_p}$ represents the contour  from 
$-\infty$ to $\infty$ and, along the way, 
passing below the pole $z_p=id/b$.
}

\vspace{5mm}
We obtained this result by the replica method. The details of this approach
and the derivation of Proposition 1 will be explained 
in Secs.~\ref{ra} and~\ref{gfhd}. In~\eqref{Result}, the factor 
$\frac{\Gamma(v_++v_-)}{\Gamma(v_++v_-+\gamma_t^{-1}d/ds)}$
comes from the Laplace transform of the pdf of $e^{\chi}$ and 
is defined by the Taylor expansion. This equation represents the relation
between ${F}_{v_{+},v_-,t}(s;X)$ and $\tilde{F}_{v_{+},v_-,t}(s;X)$, which
will be discussed particularly in Sec.~\ref{Shifting}. 

Note that the distribution function~\eqref{Pr1} is expressed as the convolution
of the Gumbel distribution and the Fredholm determinants. Such a mathematical
structure is common to the cases of the narrow wedge initial condition i.e. $h(x,0)=-|x|/\delta,~\delta\rightarrow 0$~\cite{SS2010a,SS2010b,SS2010c,ACQ2010}
and the half Brownian motion initial condition, i.e. $h(x,0)$ is BM for $x>0$
while the narrow wedge for $x<0$~\cite{CQ2010p,IS2011}. 
(This is the reason why we introduce the generalized height~\eqref{tildechi}.)
Moreover, \eqref{Pr1} can
be regarded as a generalization of the probability distributions for the
narrow wedge and half-BM initial conditions:
Noticing the relations
\begin{align}
&~\lim_{
\substack{v_+\rightarrow\infty\\ v_-\rightarrow\infty} 
}
\Ai_\Gamma^\Gamma\left(x,
\frac{1}{\gamma_t},v_+-\frac{X}{\gamma_t},v_-+\frac{X}{\gamma_t}\right)
\Ai_\Gamma^\Gamma \left(y,
\frac{1}{\gamma_t},v_-+\frac{X}{\gamma_t},v_+-\frac{X}{\gamma_t}\right)=
\Ai(x)
\Ai(y),\\
&~\lim_{
\substack{v_+\rightarrow 0\\ v_-\rightarrow\infty} 
}
\Ai_\Gamma^\Gamma\left(x,
\frac{1}{\gamma_t},v_+-\frac{X}{\gamma_t},v_-+\frac{X}{\gamma_t}\right)
\Ai_\Gamma^\Gamma \left(y,
\frac{1}{\gamma_t},v_-+\frac{X}{\gamma_t},v_+-\frac{X}{\gamma_t}\right)\notag\\
&=
\Ai_\Gamma\left(x,
\frac{1}{\gamma_t},-\frac{X}{\gamma_t}\right)
\Ai^\Gamma\left(y,
\frac{1}{\gamma_t},-\frac{X}{\gamma_t}\right),
\end{align}
where 
\begin{align}
&\Ai_\Gamma (a,b,c)=\frac{1}{2\pi}
\int_{-\infty}^{\infty} dz 
e^{iza+i\frac{z^3}{3}}\frac{1}{\Gamma\left(-ibz+c\right)},
\label{AiG2}\\
&\Ai^\Gamma (a,b,c)=\frac{1}{2\pi}
\int_{\Gamma_{i\frac{c}{b}}} dz 
e^{iza+i\frac{z^3}{3}}{\Gamma\left(ibz+c\right)},
\label{AiG1}
\end{align}
we easily find the function~\eqref{Pr1} becomes 
that for the narrow wedge case as both $v_+$ and $v_-$ go to 
infinity and that for the half Brownian motion case as $v_-$ 
($v_+$) goes to infinity (0). 

\subsubsection{Stationary limit}
Here we consider the stationary limit where both drifts 
$v_{\pm}$ in~\eqref{BMic} go to 0. 
When we take this limit for the expression~\eqref{Result}, we notice
that the Gamma function factor $\Gamma(v_++v_-)$ in~\eqref{Result}  becomes 
divergent
while the factor $\tilde{F}_{v_+,v_-,t}(s)$ vanishes 
since the term $\chi$ in~\eqref{tildechi} diverges. Thus we have to
analyze carefully the limiting behavior of each factor, which will be
discussed in Sec.~\ref{sl}.
We obtain the following result: Let us introduce the scaled drifts
$\o_{\pm}=v_{\pm}/\gamma_t$ and consider the 
quantity $F_{w,t}(s)$ defined by
\begin{align}
F_{w,t}(s;X)=\lim_{\substack{\o_-\rightarrow -\o_+\\ \o_+=w}}
F_{{\o_+}/{\gamma_t},{\o_-}/{\gamma_t},t}(s;X).
\label{defFw}
\end{align}
Note that the situation $\o_++\o_-=0$ corresponds to the case
of flat macroscopic profile (see Fig.~\ref{twosidedBM} 
and~\eqref{macroshape}) 
and the parameter $w$ determines the slope of the surface.
The case $w=0$ corresponds to the initial condition~\eqref{init_2bm}. 
In terms of the functions
\begin{align}
&B_{a,b,u}(x)
:=\frac{e^{a^3/3-(x+u)a}}{\Gamma\left(\frac{a+b}{\gamma_t}+1\right)}-
\int_{0}^{\infty}d\lambda e^{a\lambda}\Ai_{\Gamma}^{\Gamma}
\left(x+u+\lambda,\frac{1}{\gamma_t},
1+\frac{b}{\gamma_t},1+\frac{a}{\gamma_t}\right),
\label{defB}
\\
&C_t(x):=\frac{e^{\gamma_t x}}{e^{\gamma_t x}-1}.
\label{defC}
\end{align}
$F_{w,t}(s;X)$ is written as follows.

\vspace{5mm}
\noindent
{\bf Theorem 2}
{\it
\begin{equation}
F_{w,t}(s;X)
=
\frac{d/ds}{\Gamma(1+\gamma_t^{-1}d/ds)}
\int_{-\infty}^{\infty} du e^{-e^{\gamma_t(s-u)}}
\left(\nu_{w,t}(u;X)-\nu_{w,t}^{(\delta)}(u;X)\right).
\label{Result2}
\end{equation}
 Here $\nu_{w,t}(u;X)$ and $\nu_{w,t}^{(\delta)}(u;X)$ are given by
\begin{align}
&{\nu}_{w,t}(u;X)=\det\left(1-A_{w-X,-w+X}\right)
L_{w-X,-w+X}(u)+\det\left(1-A_{w-X,-w+X}-D_{w-X,-w+X}\right),
\label{nurep}
\\
&{\nu}_{w,t}^{(\delta)}(u;X)
=\det\left(1-A^{(\delta)}_{w-X,-w+X}\right)
L^{(\delta)}_{w-X,-w+X}(u)+\det\left(1-A^{(\delta)}_{w-X,-w+X}
-D^{(\delta)}_{w-X,-w+X}
\right),
\label{nudrep}
\end{align}
respectively.
The kernels $A_{a,b},~L_{a,b}$ and the function 
$D_{a,b}$ are represented as
\begin{align}
&A_{a,b}(\xi_1,\xi_2)\notag\\
&=C_t(\xi_1)
\int_{u}^{\infty}dy\Ai_{\Gamma}^{\Gamma}\left(\xi_1+y,\frac{1}{\gamma_t},
1+\frac{b}{\gamma_t},1+\frac{a}{\gamma_t}\right)
\Ai_{\Gamma}^{\Gamma}\left(\xi_2+y,\frac{1}{\gamma_t},
1+\frac{a}{\gamma_t},1+\frac{b}{\gamma_t}\right),
\label{kernela}
\\
&L_{a,b}(u)=\frac{1-a-b}{a+b}-\int_{-\infty}^{\infty}dx
C_t(x)B_{a,b,u}(x)B_{b,a,u}(x),
\label{functionl}\\
&D_{a,b}(\xi_1,\xi_2)
=\left(A_{a,b}C_tB_{a,b,u}\right)(\xi_1)B_{b,a,u}(\xi_2)
\label{kerneld}
\end{align}
where the functions $A^{(\delta)}_{a,b}$, $L^{(\delta)}_{a,b}$ and 
$D^{(\delta)}_{a,b}(u)$ are also defined 
in the same way as~\eqref{kernela},~\eqref{functionl} and~\eqref{kerneld}
respectively with $C_t(x)$~\eqref{defC} replaced by 
$C_t^{(\delta)}(x):=\frac{e^{\gamma_t x}}{e^{\gamma_t x}-1}-\delta(x)$.}

\vspace{5mm}
Although in the limit $b\rightarrow -a$, the first term in~\eqref{functionl} 
becomes divergent, we find the second term compensates for the divergence
and obtain the following expression for~\eqref{functionl},
\begin{align}
L_{a,-a}(u)
&=-\frac{2\gamma}{\gamma_t}+u-a^2-1
+\int_{-\infty}^{\infty}dxC_t(x)\left(B_{-a,a,u}^{(1)}(x)B_{a,-a,u}^{(2)}(x)
+B_{a,-a,u}^{(1)}(x)B_{-a,a,u}^{(2)}(x)\right)\notag\\
&~~-\int_{-\infty}^{\infty}dxC_t(x) 
B_{a,-a,u}^{(2)}(x)B_{-a,a,u}^{(2)}(x),
\label{lrep}
\end{align}
where $\gamma$ is Euler's constant and 
\begin{align}
&B_{a,b,u}^{(1)}(x)=\frac{e^{a^3/3-(x+u)a}}{\Gamma\left(\frac{a+b}{\gamma_t}+1\right)},
\notag
\\
&B_{a,b,u}^{(2)}(x)=\int_{0}^{\infty}d\lambda e^{a\lambda}\Ai_{\Gamma}^{\Gamma}
\left(x+u+\lambda,\frac{1}{\gamma_t},
1+\frac{b}{\gamma_t},1+\frac{a}{\gamma_t}\right)
\label{B(12)}
\end{align}
are the first and second term in~\eqref{defB}.

\begin{figure}[h]
\begin{center}
\includegraphics[scale=0.6]{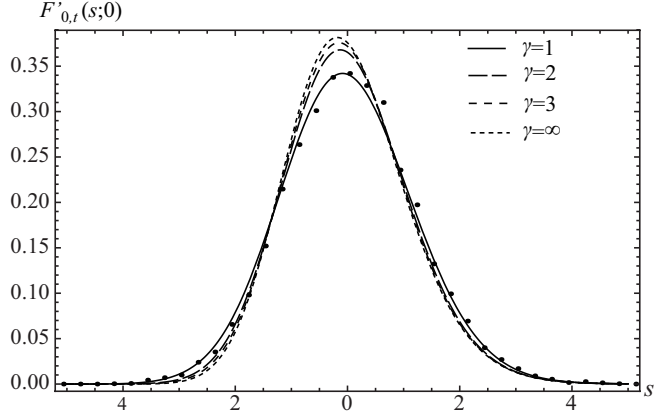}
\caption{\label{density123}
The probability density function of the height in 
the stationary situation $d F_{w=0,t}(s;X=0)/ds$ 
for $\gamma_t=1, 2, 3$. The dashed curve is $d F_0(s;0)/ds$. The dots ($\cdot$) indicate
the Monte Carlo simulation for current distribution for the WASEP.
We set the right hopping rate $q=0.575$ and the left rate $p=0.425$. 
(i.e. $\e=q-p=0.15$), $3951$ Monte Carlo steps (corresponding to $\gamma_t=1$), and 10000 samples.
}
\end{center}
\end{figure}

The expression in Theorem 2 is suitable for numerical evaluations
of the distribution function. By contrast Proposition 1 
is not because we need precise estimations for both the divergence 
of the Gamma function 
factor and the convergence of the factor
$\tilde{F}_{v_+,v_-,t}(s;X)$ to zero in the stationary limit. 
Fig.~\ref{density123}
shows the pictures of the probability density function for $w=X=0$. These are
obtained by discretizing simply the Fredholm determinants in~\eqref{nurep} 
and~\eqref{nudrep}, which is sufficient for our purpose 
of getting those pictures. A more elaborate approximation technique 
for a Fredholm determinant has been developed recently in~\cite{Bo2010,Bo20102,PS2011p}. For accurate estimations of statistical quantities of our height distributions, this technique would be useful.

The distribution function obtained in this paper appears not merely in the KPZ
equation but also in some time regime in various stochastic models~\cite{BG1997,AKQ2010,CLDR2010}. 
One typical example
is the asymmetric simple exclusion process (ASEP). The ASEP is a stochastic
many-particle system where each particle 
hops to the right (left) neighboring site with rate
$q$ ($p$) but it cannot hop to the occupied sites, 
which leads to the exclusion effect. Now we consider the 
ASEP on the one dimensional lattice $\mathbb{Z}$ and assume that
at $t=0$, particles obey the two-sided Bernoulli initial condition with density
$1/2$, where a particle occupy each site 
with probability $1/2$. Note that such a Bernoulli product measure is 
a stationary state of the ASEP.
Let $N(x,t)$ be the (integrated) particle current i.e. 
difference between the numbers of the
particles which pass from the site $x$ to the site $x+1$ and from 
the site $x+1$ to the site $x$ up to time $t$. 

In~\cite{BG1997}, it was shown that 
\begin{align}
\lim_{\e\rightarrow 0}\sqrt{\e}\left(2N(\e^{-1}x,\e^{-2}t)+t\e^{-3/2}/2\right)
=h_{\frac{1}{2},1,1}(x,t)+\frac{t}{24}
\end{align}
where $\e:=q-p$ is the difference between the right ($q$) and the left ($p$)
hopping rates. The function $h_{\frac{1}{2},1,1}(x,t)$ 
is the (Cole-Hopf) solution to the KPZ equation 
with parameters $\nu=1/2,~\lambda=1,~D=1$ and with the initial condition 
given in~\eqref{BMic} with $v_+=v_-=0$.
This relation indicates that the height of the KPZ equation
well approximates the current in the ASEP  
with the asymmetry $q-p$ being small, which is called 
the weakly ASEP (WASEP). Moreover it has been known recently that
the KPZ equation also describes the dynamics of many other weakly driven 
growth processes in some time regime~\cite{AKQ2010,CLDR2010}.
The dots in Fig.~\ref{density123} 
represent the Monte Carlo simulation for the particle current fluctuation
$\sqrt{\e}\left(2N(0,\e^{-2}t)+t\e^{-3/2}/2\right)$
for the WASEP with the two-sided Bernoulli initial condition mentioned above.

\subsubsection{Long-time limit}
We can also obtain the long-time limit of the height distribution function.
We define $F_w(s;X)$ as $\lim_{t\rightarrow\infty}F_{w,t}(s;X)$ 
and obtain the following result.

\vspace{5mm}
\noindent
{\bf Theorem 3}
{\it
\begin{equation}
F_{w}(s;X)
=
\frac{d}{ds}\nu_w(s;X)
.
\label{Result4}
\end{equation}
where $\nu_w(s;X):=\lim_{t\rightarrow\infty}\nu_{w,t}(s;X)$
has the following expression.
\begin{align}
\nu_w(s;X)=\det\left(1-\A\right)\L_{w-X,-w+X}(s)
+\det\left(1-\A-\D_{w-X,-w+X}\right).
\label{nuw}
\end{align}
Here $\A$, $\D_{a,b}$ and $\L_{a,b}$ are the long-time limit of $A_{a,b}$, $D_{a,b}$ and $L_{a,b}$ defined in Theorem 2 and are given by
\begin{align}
&\A(x,y)=P_0(x)\int_{u}^{\infty}d\xi\Ai(x+\xi)\Ai(y+\xi),
\label{mcA}
\\
&\L_{a,b}(s)=\frac{1-a-b}{a+b}-\int_0^{\infty}dx \B_{a,s}(x)\B_{b,s}(x),
\label{mcLw}
\\
&\D_{a,b}(x,y)=\left(\A P_0 \B_{a,s}\right)(x)\B_{b,s}(y),
\label{mcDw}
\\
&\B_{a,s}(x)=e^{\frac{a^3}{3}-(x+s)a}-\int_{0}^{\infty}d\lambda e^{a\lambda}
\Ai(x+s+\lambda).
\label{mcBw}
\end{align}
}

\vspace{5mm}
Note that $\det\left(1-\A\right)$ in~\eqref{nuw} is nothing but the GUE
Tracy-Widom distribution. The expression for $\L_{a,-a}$ 
corresponding to~\eqref{lrep} is given by
\begin{align}
&\L_{a,-a}(s)=s-a^2-1+\int_s^{\infty}dx\left(\B_{-a,s}^{(1)}(x)\B_{a,s}^{(2)}(x)
+\B_{a,s}^{(1)}(x)\B_{-a,s}^{(2)}(x)\right)-\int_{s}^{\infty}dx 
\B_{a,s}^{(2)}(x)\B_{-a,s}^{(2)}(x),
\end{align}
where $\B_{a,s}^{(1)}(x)$ ($\B_{a,s}^{(2)}(x)$) represents the first (second)
term in~\eqref{mcBw}, i.e.
\begin{align}
\B_{a,s}^{(1)}(x)=e^{\frac{a^3}{3}-(x+s)a},~~
\B_{a,s}^{(2)}(x)=\int_{0}^{\infty}d\lambda e^{a\lambda}
\Ai(x+s+\lambda).
\end{align}

This function $F_w(s)$ has already appeared as a limiting distribution 
for other stochastic processes such as the PNG
model~\cite{BR2000} and the TASEP~\cite{FS2006} 
in stationary situation. 
This implies that $F_w(s;X)$ appears commonly in the stationary subclass 
in the KPZ universality. Our result above is a representation which
does not include any resolvent kernel such as $(1-\A)^{-1}(x,y)$
thus it is convenient for numerical calculation.
In Fig.~\ref{density123}, we also depicted the picture for 
the long-time limit using~\eqref{Result2}.

The derivation of the above theorem will be provided
in Sec.~\ref{lmt}.

\subsection{The two-point correlation functions}
The two-point function of the height is defined as
\begin{align}
C(x,t)=\left\langle
(h(x,t)-\langle h(x,t)\rangle)^2
\right\rangle.
\label{cxt}
\end{align}
Considering the scaling property~\eqref{scaled}, we also introduce 
the scaling form
\begin{align}
g_t(y)=2^{-4/3}\left(\frac{\lambda}{2\nu\gamma_t}\right)^2
C\left(2^{1/3}\frac{2\gamma_t^2}{\a^2}y,t\right)
\label{s2f}
\end{align}
where we put the overall factor $2^{-4/3}$ and the factor $2^{1/3}$
in the argument of $C(x,t)$ following the convention in~\cite{PS2004}.

Moreover the two-point function of the slope of the height 
$\left\langle
\partial_xh(0,0)\partial_xh(x,t)
\right\rangle$
is also an important quantity since the height slope $\partial_xh(x,t)$
corresponds to the particle density in the interacting particle processes
such as the ASEP. It has been pointed out in~\cite{PS2004} that 
this is expressed as 
the second derivative of $C(x,t)$,
\begin{align}
\left\langle
\partial_xh(0,0)\partial_xh(x,t)
\right\rangle=\frac12\partial_x^2C(x,t)
\end{align}
Thus we are also interested in the second derivative of the scaling 
function $g''_t(y)$.

Noting that~\eqref{cxt} is just the variance of $h(x,t)$, 
we readily obtain the following result for $g_t(y)$:

\vspace{5mm}
\noindent
{\bf Corollary 4}
{\it
\begin{align}
g_t(y)=2^{-4/3}\left(
\int_{-\infty}^{\infty}s^2\frac{dF_{w=0,t}(s;2^{1/3}y)}{ds}ds
-\left(
\int_{-\infty}^{\infty}s\frac{dF_{w=0,t}(s;2^{1/3}y)}{ds}ds\right)^2\right)
\end{align}
where $F_{w,t}(s;X)$ is given in~\eqref{Result2}.
}

\begin{figure}[h]
\begin{center}
\includegraphics[scale=.6]{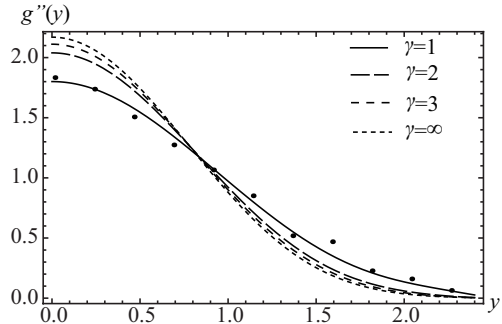}
\caption{\label{twopt123} Stationary two-point correlation function $g_t''(y)$ for the 
KPZ equation for $\gamma_t=1,2,3$. The dashed curve represents the one 
in the scaling limit $g''(y)$. The dots indicates the Monte Carlo result of the WASEP, where parameters are the same as those in Fig. 2.}
\end{center}
\end{figure}

\vspace{5mm}
Fig.~\ref{twopt123} shows the function $g''_t(y)$ for $\gamma_t=1,2,3$. 
The dots in this figure correspond to the Monte Carlo simulation of 
density correlation in the WASEP
\begin{align}
\langle \eta_{x,t}\eta_{0,0}\rangle_{\text{WASEP}}-1/4
\end{align}
where $\eta_{x,t}$ is the occupation variable. i.e. if the site $x$ is occupied
 (empty) at time $t$, $\eta_{x,t}=1~(0)$ and $\langle~~\rangle_{\text{WASEP}}$
means the averaging over both two-sided Bernoulli initial configuration and 
WASEP dynamics.

\section{Replica analysis}\label{ra}

\subsection{Cole-Hopf solution, $\delta$-Bose gas}
By the scaling of the KPZ equation, we find the height 
$h_{\nu,\lambda,D}(x,t)$ of the KPZ
equation with general parameters $\nu,~\lambda,~D$  and 
$h_{\frac12,1,1}(x,t)$
with the specialized parameters $\nu = \frac12, \lambda = D=1,  $ 
are related as
\begin{align}
\frac{\lambda}{2\nu}h_{\nu,\lambda,D}\left(x,\frac{t}{2\nu}\right)
=h_{\frac12,1,1}(\a^2x,\a^4t),
\label{generalspecific}
\end{align}
where $\a$ is defined below~\eqref{init_2bm},
for details see~\cite{IS2011}.
Thus in the following we concentrate on the height $h_{\frac12,1,1}(x,t)$
and omit the indices $\frac12,1,1$.

Applying the Cole-Hopf transformation, 
\begin{equation}
 Z(x,t) = e^{h(x,t)},
\label{zdef}
\end{equation}
the KPZ equation (\ref{KPZ}) is linearized as 
\begin{equation}
\frac{\partial Z(x,t)}{\partial t}=
\frac12 \frac{\partial^2 Z(x,t)}
{\partial x^2}+\eta(x,t)Z(x,t),
\label{SHE}
\end{equation}
which is called the stochastic heat equation. 
Now we would like to remark on the regularization of the KPZ equation. 
The KPZ equation~\eqref{KPZ} 
is in fact not well-defined due to the interplay between the nonlinear term 
and the singular property of the noise in~\eqref{KPZ}. 
The typical regularization is to interpret \eqref{SHE} as It{\^o} type and
to define the regularized KPZ equation by the inverse Cole-Hopf transformation 
of~\eqref{SHE}.
This is called the Cole-Hopf solution of the KPZ equation~\cite{BG1997}. 
For more detailed discussion, see~\cite{IS2011}. Recently a new way 
of regularization without using the Cole-Hopf transformation~\eqref{zdef}
was proposed in~\cite{H2011p}.

Eq.~\eqref{SHE} can be interpreted as a problem of directed polymer in 
random media. Applying the Feynman path integral representation 
and taking the average with respect to $\eta$, we see 
the problem becomes that of 
the delta function Bose gas with attractive interaction, 
\begin{equation}
H_N=-\frac{1}{2}\sum_{j=1}^N\frac{\partial^2}{\partial x_j^2}
-\frac12\sum_{j\neq k}^N\delta(x_j-x_k),
\label{dBose}
\end{equation}
in terms of which the replica partition function 
$\langle Z^N(x,t)\rangle$ can be written as
\begin{equation}
\langle Z^N(x,t) \rangle=\langle x|e^{-H_Nt}|\Phi\rangle.
\label{KPZdBose1}
\end{equation}
Here $\langle x|$ represents the state with all $N$ particles 
being at the position $x$ and the $|\Phi\rangle$ the initial 
state of the $\delta$-function Bose gas. This is an 
important property of the replica approach to 
the KPZ equation~\cite{CLDR2010,K1987,D2010}. 
The integrability of the Hamiltonian~\eqref{dBose} allows us to
calculate exactly the replica partition function.

One can perform 
the average over the initial distribution given by the two-sided 
Brownian motion and the dependence 
of $|\Phi\rangle$ on $x_1,\ldots,x_N$ can be explicitly calculated as
follows. First let us assume $x_1 < x_2 < \ldots < x_N$. 
For $x_1 < \ldots < x_l < 0 < x_{l+1}<\ldots <x_N, 1\leq l \leq N$,
one finds 
\begin{align}
 &\quad \langle x_1, \cdots, x_N|\Phi\rangle \notag\\
 &= 
 \langle \exp\left[ B_-(-x_1)+\cdots +B_-(-x_l) 
                    +B_+(x_{l+1})+\cdots +B_+(x_N) \right]\rangle \notag\\
 &=
 e^{v_-\sum_{j=1}^l x_j -v_+\sum_{j=l+1}^N x_j}
 \prod_{j=1}^l e^{-\frac12(2l-2j+1)x_l}
 \prod_{k=1}^{N-l} e^{\frac12(2(N-l-k)+1)x_{l+k}},
\end{align}
where $\langle~~\rangle$ denotes the averaging over 
BMs $B_{\pm}(x)$~\eqref{BMic} 
and in the last equality we used the property of BM, 
$\langle B(x)B(y) \rangle=\min(x,y)$.
Since we are considering a Boson system, 
we symmetrize the function with respect to $x_1,\ldots,x_N$ to have
\begin{align}
\langle x_1,\cdots,x_N|\Phi\rangle
&=\sum_{P\in S_N}
 e^{v_-\sum_{j=1}^l x_{P(j)} -v_+\sum_{j=l+1} x_{P(j)}}
\prod_{j=1}^l e^{-\frac12(2j-1)x_{P(j)}}\notag\\
&~~\times\prod_{k=1}^{N-l} e^{\frac12(2(N-l-k)+1)x_{P(l+k)}}
\prod_{j=1}^{N}\Theta\left(x_{P(j)}-x_{P(j-1)}\right).
\label{halfBMic}
\end{align}
Here $S_N$ denotes the set of permutations of order $N$,
$\Theta(x)$ is the step function and we set $x_{P(0)}=0$.

The eigenvalues and eigenfunctions of the $\delta$-Bose gas
can be constructed by using the Bethe 
ansatz~\cite{LL1963,M1964} (see also~\cite{D2010}). 
Let $|\Psi_z\rangle$ and $E_z$ be the eigenstate and its eigenvalue
of $H_N$,
\begin{equation}
H_N|\Psi_z\rangle=E_z|\Psi_z\rangle.
\label{esev}
\end{equation}
By the Bethe ansatz, the eigenstate is given as
\begin{align}
\langle x_1,\cdots, x_N|\Psi_z\rangle
=C_z\sum_{P\in S_N}{\text{sgn}P}\prod_{1\le j<k\le N}
\left(z_{P(j)}-z_{P(k)}+i\text{sgn}(x_j-x_k)\right) \exp\left(i\sum_{l=1}^N
z_{P(l)}x_l\right)
\label{eigenfunction}
\end{align}
where $C_z$ is the normalization constant, for which a formula 
is given in (\ref{norma}) below. 

For the $\delta$-Bose gas with attractive interaction,
the quasimomenta $z_j~(1\le j\le N)$, which label the state,
are in general complex numbers. $z_j~(1\le j\le N)$ are divided 
into $M$ groups where $1\leq M \leq N$. 
The $\a$th group consists of $n_\a$ quasimomenta
$z_j's$ which share the common real part $q_{\a}$.
Note that 
$\sum_{\a=1}^Mn_\a=N$. 
The quasimomenta in each group line up with
regular intervals with unit length 
along the imaginary direction. 
Using $q_{\a}$ and $n_{\a}~(1\le \a\le M)$, we represent 
$z_j$ $(1\le j\le N)$ as
\begin{align}
z_j=q_{\a}-\frac{i}{2}\left(n_{\a}+1-2r_\a\right),~~\text{for}~
j=\sum_{\b=1}^{\a-1}n_\b+r_\a,
\label{zq}
\end{align}
where $1\le \a\le M$ and $1\le r_\a \le n_\a$.
The normalization constant $C_z$ in~\eqref{eigenfunction}, 
which is taken to be a positive real 
number, and the eigenvalue $E_z$ are given by~\cite{D2010}
\begin{align}
C_z
&=\left( \frac{1}{ N!}
\prod_{\a=1}^M\frac{(n_\a!)^2}{n_\a} 
\prod_{1\le j<k\le N}
\frac{1}{|z_j-z_k-i|^2} \right)^{1/2},
\label{norma}\\
E_z&=\frac12\sum_{j=1}^N z_j^2=\frac{1}{2}\sum_{\a=1}^Mn_\a q_\a^2
-\frac{1}{24}\sum_{\a=1}^M \left(n_\a^3-n_\a\right).
\label{eigenvalue}
\end{align}
The completeness of these states has been proved 
in~\cite{O1979th,HO1997,PS2011p2}.

The replica partition function 
$\langle Z^N(x,t)\rangle$~\eqref{KPZdBose1} can be written as
\begin{align}
\langle Z^N(x,t) \rangle&=\int_{-\infty}^{\infty}dy_1\cdots
\int_{-\infty}^{\infty}dy_N
\langle x|e^{-H_Nt}|y_1,\cdots,y_N\rangle\langle y_1,\cdots,y_N|\Phi\rangle
\label{KPZdBose}
\end{align}
Expanding the propagator $\langle x|e^{-H_Nt}|y_1,\cdots,y_N\rangle$
by the Bethe eigenstates of the $\delta$-Bose gas
(\ref{eigenfunction}), we have
\begin{align}
\langle Z^N(x,t)\rangle
&=
\sum_{M=1}^N\frac{1}{M!}
\prod_{j=1}^N\int_{-\infty}^{\infty}dy_{j}
\left(\int_{-\infty}^\infty 
\prod_{\a=1}^M\frac{dq_\a}{2\pi}\sum_{n_\a=1}^{\infty} \right)
\delta_{\sum_{\beta=1}^M n_\beta,N} \notag\\
&\quad \times
e^{-E_zt}\langle x|\Psi_z\rangle\langle\Psi_z|y_1,\cdots,y_N\rangle
\langle y_1,\cdots,y_N|\Phi\rangle .
\label{KPZdBose2}
\end{align}
Now we want to perform the integrations over $y_j,~(1\le j\le N)$ 
in this equation,
Using~\eqref{halfBMic} and 
noticing the symmetry of the eigenfunction~\eqref{eigenfunction}, 
$\langle\Psi_z|y_{P(1)},\cdots,y_{P(N)}\rangle
=\langle\Psi_z|y_{1},\cdots,y_{N}\rangle$, we find that 
RHS of~\eqref{KPZdBose2} is represented as
\begin{align}
\langle {Z}^N(x,t)\rangle
&=\sum_{M=1}^N\frac{N!}{M!}\sum_{\ell=0}^N
\int_{y_1<\cdots<y_{\ell}<0<y_{\ell+1}<\cdots <y_N}dy_1\cdots dy_N
\prod_{\a=1}^M\left(
\int_{-\infty}^\infty\frac{dq_\a}{2\pi}
\sum_{n_\a=1}^{\infty}\right)\delta_{\sum_{\beta=1}^Mn_\beta,N}
\notag\\
&~\times e^{-E_zt}
\langle x|\Psi_z\rangle\langle\Psi_z|y_1,\cdots,y_N\rangle
\notag\\
&\times
 e^{v_-\sum_{j=1}^l y_j -v_+\sum_{j=l+1} y_j}
 \prod_{j=1}^l e^{-\frac12(2j-1)y_j}
 \prod_{k=1}^{N-l} e^{\frac12(2(N-l-k)+1)y_{l+k}}.
\label{Betheexpand}
\end{align}
Here we find the necessity of the drift terms in~\eqref{BMic}. 
Note that due to the factors
$\exp(-\frac12(2l-2j+1)y_j)$ and $\exp(\frac12(2(N-l-k)+1)$ we could not perform the integrations over $y_k,~k=1,\cdots,N$
prior to those over $q_\a,~\a=1,\cdots,M$ if we did not introduce
the drifts $v_{\pm}$. 
Now we perform the integrations of $y_j~(1\le j \le N)$ by assuming 
that the drifts are positively large enough, which is the main 
reason why we introduced the drifts.
Eq.~\eqref{KPZdBose2} can now be expressed as
\begin{align}
\langle {Z}^N(x,t)\rangle
=\sum_{M=1}^N\frac{1}{M!}
\prod_{\a=1}^M\left(
\int_{-\infty}^{\infty}\frac{dq_\a}{2\pi}
\sum_{n_\a=1}^{\infty}\right)\delta_{\sum_{\beta=1}^Mn_\beta,N}
\langle x|\Psi_z\rangle\langle\Psi_z|\Phi\rangle
e^{-E_zt}.
\label{afterdeform}
\end{align}
Here $\langle x|\Psi_z\rangle$ is given by 
\eqref{eigenfunction} with $x_1=\cdots=x_N=x$ and
$\langle\Psi_z|\Phi\rangle$ is computed as 
\begin{align}
&~\langle \Psi_z|\Phi\rangle
=\prod_{j=1}^N\int_{-\infty}^{\infty}dy_j\langle\Psi_z|y_1,\cdots,y_N\rangle
\langle y_1,\cdots,y_N|\Phi\rangle
\notag\\
&=N! C_z
\sum_{P\in S_N}\text{sgn}P
\left(z^*_{P(j)}-z^*_{P(k)}+i)\right)
\sum_{l=0}^N \int_{y_1<\cdots < y_l<0} dy_1\cdots dy_l
\prod_{m=1}^l e^{\left(-i z^*_{P(m)}+v_--\frac12(2m-1)\right)y_m}
\notag\\&\quad \times
\int_{0<y_{l+1}<\cdots < y_N} dy_{l+1}\cdots dy_N  
\prod_{m=1}^{N-l} 
e^{\left(-i z^*_{P(l+m)}-v_+ +\frac12(2(N-l-m)+1)\right)y_{l+m}}
\notag\\
&=(-1)^NN!C_z \sum_{P\in S_N}{\text{sgn}P}\prod_{1\le j<k\le N}
\left(z^*_{P(j)}-z^*_{P(k)}+i\right) 
\notag\\&\quad \times
\sum_{l=0}^N (-1)^l 
\prod_{m=1}^l \frac{1}{\sum_{j=1}^m (-iz^*_{P{(j)}}+v_-)-m^2/2}
\prod_{m=1}^{N-l} \frac{1}{\sum_{j=N-m+1}^N (-iz^*_{P{(j)}}-v_+)+m^2/2}.
\label{nodangerous}
\end{align}
Here we assume that $\sum_{j=1}^m (-iz^*_{P{(j)}}+v_-)-m^2/2>0$ and
$\sum_{j=N-m+1}^N (-iz^*_{P{(j)}}-v_+)+m^2/2$ for any $m~(1\le m\le N)$.
For example if we set $v_{\pm} >N/2+\max_{\a}n_{\a}/2$, the above conditions
are satisfied.

\subsection{Combinatorial identities}
For further analysis of the integrand in~\eqref{afterdeform},
we need two combinatorial identities for $\langle x|\Psi_z\rangle$ and
$\langle\Psi_z|\Phi\rangle$. The first one is for
$\langle x|\Psi_z\rangle$. One has 
\begin{equation}
\sum_{P\in S_N}{\text{sgn}P}\prod_{1\le j<k \le N}
\left(w_{P(j)}-w_{P(k)}+if(j,k)\right)
=N!\prod_{1\le j<k\le N}(w_j-w_k)
\label{ci1}
\end{equation}
for any complex variables $w_j$~$(1\le j \le N)$ and 
$f(j,k)$. This identity was derived as Lemma~1 in~\cite{PS2010p}.

The next one for the term $\langle\Psi_z|\Phi\rangle$ is 

\vspace{3mm}
\noindent
{\bf Lemma 5}
{\it 
For any complex numbers $w_j~(1\le j\le N)$ and $a$,  
\begin{align}
&~\sum_{P\in S_N}{\text{\rm sgn} P}\prod_{1\le j<k \le N}
\left(w_{P(j)}-w_{P(k)}+a\right) 
\notag\\&\quad \times
\sum_{l=0}^N (-1)^l 
\prod_{m=1}^l \frac{1}{\sum_{j=1}^m (w_{P(j)}+v_-)-m^2a/2}
\prod_{m=1}^{N-l} \frac{1}{\sum_{j=N-m+1}^N (w_{P{(j)}}-v_+)+m^2a/2}
\notag\\
&~=
\frac{ \prod_{m=1}^N (v_++v_--a m)  \prod_{1\le j<k\le N}(w_j-w_k)}
     { \prod_{m=1}^N(w_m+v_--a/2) (w_m-v_++a/2)}.
\label{ci2}
\end{align}
}

\vspace{3mm}
\noindent
{\bf Proof} 

First we note that a similar identity, 
\begin{align}
\sum_{P\in S_N}{\text{\rm sgn} P}\prod_{1\le j<k \le N}
\left(w_{P(j)}-w_{P(k)}+a\right) 
\prod_{m=1}^{N-l} \frac{1}{\sum_{j=N-m+1}^N w_{P(j)}+m^2a/2}
=
\frac{ \prod_{1\le j<k\le N}(w_j-w_k)}{ \prod_{m=1}^N(w_m+a/2) },
\label{cihb}
\end{align}
which is the special case $v_-\rightarrow -\infty,~v_+\rightarrow 0$ 
of the above (\ref{ci2}), was already proved 
in~\cite{IS2011}. 
By using this identity for both terms with $v_+$ and $v_-$ on LHS, 
one can rewrite LHS of (\ref{ci2}) as 
\begin{gather}
 \sum_{n=0}^N (-1)^n 
\sum_{\substack{1\leq j_1 < \ldots <j_n\leq N \\ 1\leq k_1 < \ldots <k_{N-n}\leq N \\
            \{j_l\}_{l=1}^n\cap\{k_m\}_{m=1}^{N-n} = \phi} }
 \text{sgn}S
 \prod_{1\le l < l'\le n} (w_{j_l}-w_{j_{l'}})  
 \prod_{1\le m < m'\le N-n} (w_{k_m}-w_{k_{m'}})\notag\\  
 \times\prod_{l=1}^n\prod_{m=1}^{N-n} (w_{j_l}-w_{k_m}+a)  
  \prod_{l=1}^n \frac{1}{w_{j_l}+v_--a/2} 
           \prod_{m=1}^{N-n} \frac{1}{w_{k_m}-v_++a/2}.  
\end{gather}
where $S=\{j_1,\cdots,j_n,k_1,\cdots,k_{N-n}\}$.
At this point we multiply both sides by $\prod_{m=1}^N(w_m+v_--a/2) (w_m-v_++a/2)$ 
and then set $v_+=a/2,v_-=b+a/2$ without losing the generality.  
Hence what we want to show is 
\begin{align}
&\sum_{n=0}^N (-1)^n 
\sum_{\substack{1\leq j_1 < \ldots <j_n\leq N \\ 1\leq k_1 < \ldots <k_{N-n}\leq N \\
            \{j_l\}_{l=1}^n\cap\{k_m\}_{m=1}^{N-n} = \phi} }
 \text{sgn}S
 \prod_{1\le l < l'\le n} (w_{j_l}-w_{j_{l'}})  
 \prod_{1\le m < m'\le N-n} (w_{k_m}-w_{k_{m'}})\notag\\  
 &\times\prod_{l=1}^n\prod_{m=1}^{N-n} (w_{j_l}-w_{k_m}+a)  
 \prod_{l=1}^n w_{j_l}  \prod_{m=1}^{N-n} (w_{k_m}+b)
           =
        \prod_{m=0}^{N-1} (b-a m)  \prod_{1\le j<k\le N}(w_j-w_k).
        \label{ciab}
\end{align}

Here we prove this by noting that both sides can be considered as a polynomial 
in $a$ and $b$ and arguing by induction that the coefficients (which are 
polynomials of
$\{w_1,\ldots,w_N\}$) for each $b^na^m$ in both sides are the same. 
First we observe that LHS is anti-symmetric 
with respect to $\{w_1,\ldots, w_N\}$. Hence 
the coefficient $b^na^m$ is also anti-symmetric 
with respect to $\{w_1,\ldots, w_N\}$.
Then all we should check is the numerical coefficients of terms 
$\prod_{j=1}^N w_j^{c_j}$ satisfying 
$c_N>\cdots>c_1\ge0$.
For instance when $N=2$, the equality we want to show reads 
\begin{align}
   &(w_1-w_2)(w_1+b)(w_2+b) - (w_1-w_2+a) w_1 (w_2+b)
 +(w_2-w_1+a)(w_1+b)w_2 \notag\\
  &+ (w_1-w_2) w_1 w_2 = b(b-a) (w_1-w_2) .
\end{align}
We compare the coefficients of $1,a,b,ba,b^2$. For example for 
that for $b$, we see the coefficient of $w_1w_2^2$ is zero;  
for $ab$, we see the coefficient of $w_2$ is one. 

\begin{figure}
\begin{picture}(500,500)
\put(0,0){\includegraphics[width=300pt]{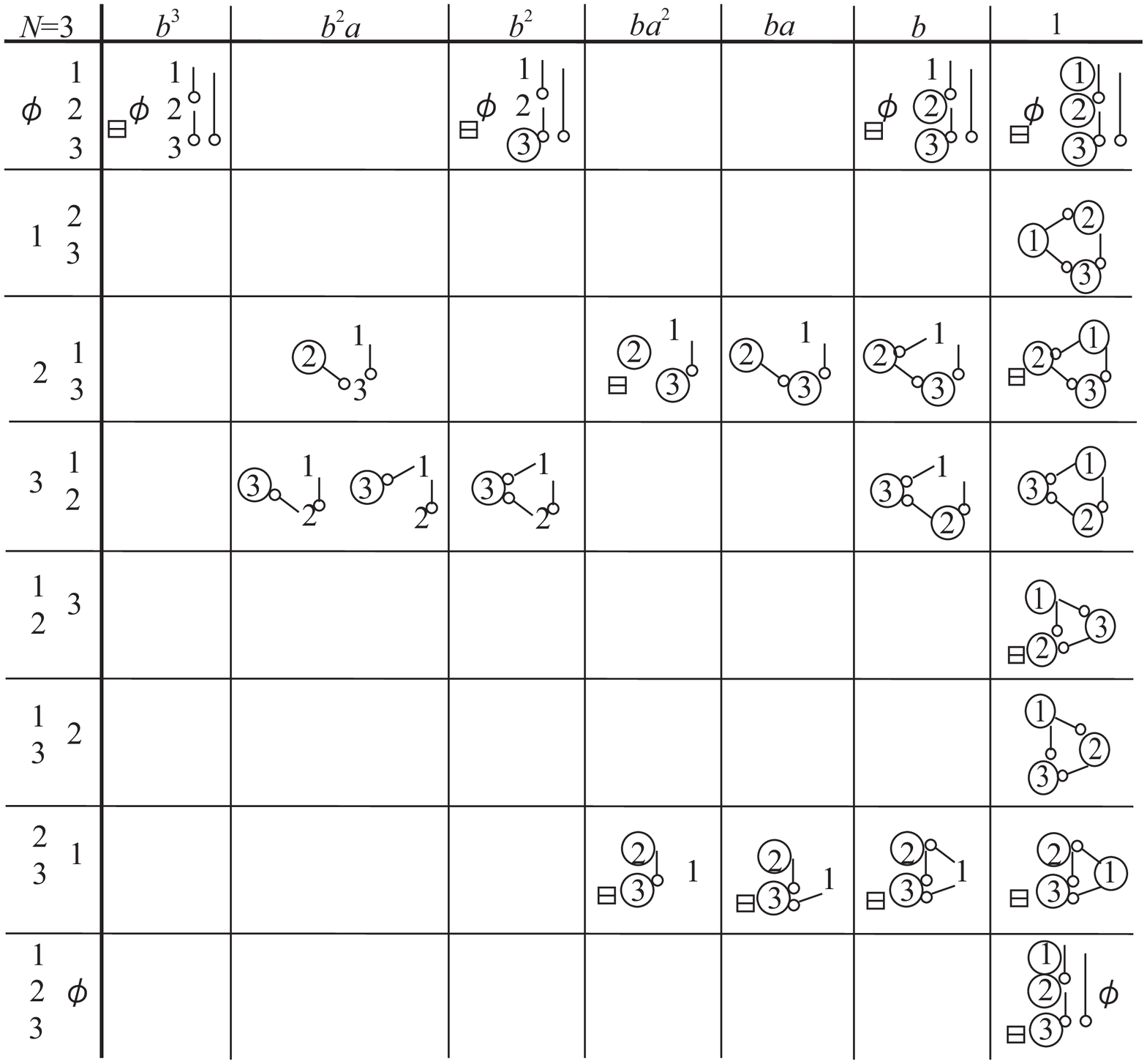}}
\put(0,310){\includegraphics[width=400pt]{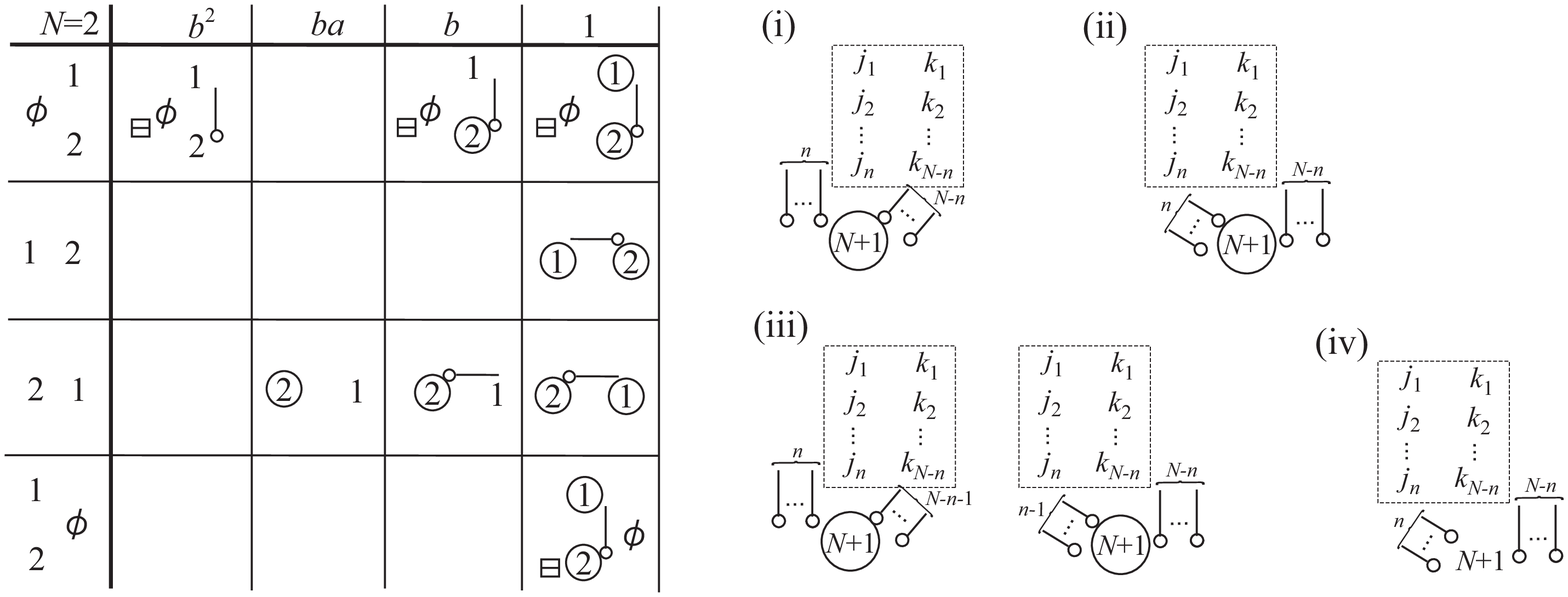}}
\put(0,485){(a)}
\put(200,485){(c)}
\put(0,290){(b)}
\end{picture}
\caption{\label{gr}Graphical representations for the coefficient of each term
in LHS of~\eqref{ciab}. (a) and (b) represents the case $N=2$ 
and $N=3$ respectively. In (c), we depict the possible four types of graph at $N-1$
which comes from the graph at $N$ 
corresponding to RHS of~\eqref{ciab}.
}
\end{figure}

To consider general $N$, we introduce a graphical representation. 
Each graph corresponds to each term in LHS of (\ref{ciab}). 
First we put 
$n$ numbers $\{j_1,\ldots ,j_n\}$ to the left and $N-n$ numbers 
$\{k_1,\ldots,k_{N-n} \}$ to the right. They are the vertices in a graph.
Next if the term contains a factor from 
$w_i-w_j$ we draw a line joining $i$ and $j$ (edge); we also put circle 
on $i$ if the term contains $w_i$ and $j$ if $w_j$. In addition, if the 
term contains a factor coming from the last two products in LHS 
of (\ref{ciab}), 
we encircle the number; hence we encircle all $j_i$'s. Finally by considering 
how each term can appear, we can put the sign to each graph. Because of 
the antisymmetry with respect to $\{w_1,\cdots,w_N\}$, 
we need to list up only the graphs with $c_N>\cdots>c_1\ge 0$. 
The example of these graphs for 
$N=2,3$ are given in Figs.~\ref{gr} (a) and (b) respectively. 
By looking at these graphs,  one can 
confirm that there are several graphs which give non-zero contributions and 
all others give zero. 

For showing the identity for general $N$, we use an induction.
Suppose that we already showed the identity for $N$ by using the graphical 
representation. It consists of the several graphs corresponding to 
RHS of the identity (see e.g. the columns ``$b^3$", ``$b^2a$" and ``$ba^2$" 
in Fig.~\ref{gr} (b)), where the corresponding coefficient is 
$\prod_{i=1}^Nw_i^{i-1}$ up to sign, 
and all other graphs being grouped to vanish.
Next we want to prove the identity for $N+1$. 
First notice that every graph which appears for $N+1$ is obtained by adding 
the vertex $N+1$ to some graph for $N$.  If it is from a graph which can be 
grouped to vanish at $N$, then the term at $N+1$ can also be grouped to vanish. 
On the other hand, by adding $N+1$ to 
a graph corresponding to RHS of the identity at $N$, 
only four possible types of graphs at $N+1$ 
corresponding to the coefficient $\prod_{j=1}^{N+1}w_{j}^{c_{j}}$ 
with $c_{N+1}>c_{N}(=N-1)>c_{N-1}(=N-2)>\cdots>c_2(=1)>c_1(=0)$ 
can be produced. 
Fig.~\ref{gr} (c) illustrates these four types. If we suppose the base graph 
at $N$ represents the coefficient $\prod_{j=2}^{N}w_j^{j-1}$ of $b^na^{N-n}$,
the types (i) and (ii) at $N+1$ (produced by adding the vertex
$``N+1"$ to the base one) 
correspond to the coefficient $(-1)^{N+1}w^{N+1}\prod_{j=2}^{N}w_j^{j-1}$ and 
$(-1)^Nw^{N+1}\prod_{j=2}^{N}w_j^{j-1}$ of $b^na^{N-n}$ respectively, 
which cancel out each other. The types (iii) and (iv), on the other hand,
correspond to the coefficient
$(-1)^{N+1}\prod_{j=2}^{N+1}w_j^{j-1}$ of $b^{n}a^{N+1-n}$ and  
$(-1)^{N}\prod_{j=2}^{N+1}w_j^{j-1}$
of $b^{n+1}a^{N-n}$ respectively. Note also that in the type (iii),
there are $N$ choices in total for connecting the vertex $N+1$ and 
the other ones with lines.
Thus we find that for $m=0,1,\cdots, N$
\begin{align}
C[b^{N+1-m}a^m]=(-1)^Nw_{N+1}^{N}C[b^{N-m}a^m]+(-1)^{N-1}Nw_{N+1}^N
C[b^{N-(m-1)}a^{m-1}],
\label{Coefba}
\end{align}
where $C[b^{n}a^m]$ represents the coefficient of $b^{n}a^m$.
This completes the proof. 
\qed

\vspace{5mm}
Using the identities~\eqref{ci1} and~\eqref{ci2} to~\eqref{eigenfunction} 
and~\eqref{nodangerous} respectively, 
we get
\begin{align}
&\langle x|\Psi_z\rangle
=N!C_z\prod_{1\le j<k\le N}(z_j-z_k)e^{i\sum_{l=1}^Nz_lx},
\label{Psicompact}
\\
&\langle\Psi_z|\Phi\rangle
=
N!C_z\prod_{1\le j<k\le N}(z^*_j-z^*_k)
\prod_{l=1}^N\frac{v_++v_- -l}{(z^*_l+iv_- -i/2)(z^*_l-iv_+ +i/2)}.
\end{align} 
Thus using these relations and~\eqref{norma}, we find that the factor
$\langle x|\Psi_z\rangle\langle\Psi_z|\Phi\rangle$
in~\eqref{afterdeform} becomes
\begin{align}
\langle x|\Psi_z\rangle\langle\Psi_z|\Phi\rangle
=N!\prod_{\a=1}^M\frac{(n_{\a}!)^2}{n_{\a}}
\prod_{1\le j<k\le N}\frac{|z_j-z_k|^2}{|z_j-z_k-i|^2}
\prod_{l=1}^N\frac{e^{iz_lx}(v_++v_--l)}{(-iz^*_l+v_--1/2)(iz_j^*+v_+-1/2)},
\label{integrand0}
\end{align}

We want to rewrite this equation in terms of $q_{\a}$ and $n_{\a}$ 
in~\eqref{zq}.
For the last factor of this equation, we easily find 
\begin{align}
&\quad\prod_{l=1}^N\frac{e^{iz_lx}}{ (-iz^*_l+v_--1/2)(iz_j^*+v_+-1/2) } \notag\\
&=
\prod_{\a=1}^Me^{in_\a q_\a x}
\prod_{r=1}^{n_{\a}}
\frac{1}{(-iq_{\a}+v_-+\frac12(n_{\a}-2r))(iq_{\a}+v_++\frac12(n_{\a}-2r))}.
\end{align} 
From~\eqref{norma} and~\eqref{Psicompact}, we know the remaining factors 
in~\eqref{integrand0} is represented by $|\langle 0|\Psi_z\rangle|^2$.
For this quantity, the following result was obtained in Appendix B 
in~\cite{D2010},
\begin{align}
|\langle 0|\Psi_z\rangle|^2&=N!\prod_{\a=1}^M\frac{(n_{\a}!)^2}{n_{\a}}
\prod_{1\le j<k\le N}\frac{|z_j-z_k|^2}{|z_j-z_k-i|^2}\notag\\
&=\frac{N!}{\prod_{\a=1}^Mn_{\a}}
\prod_{1\le\a<\b\le M}\frac{|q_\a-q_\b-\frac{i}{2}(n_{\a}-n_{\b})|^2}
{|q_\a-q_\b-\frac{i}{2}(n_{\a}+n_{\b})|^2}.
\label{Psi0}
\end{align}
From~\eqref{integrand0}--\eqref{Psi0}, we get
\begin{align}
\langle x|\Psi_z\rangle\langle\Psi_z|\Phi\rangle
&=
N!\prod_{\ell=1}^N(v_++v_--\ell)
\prod_{\a<\b}^M
\frac{|q_\a-q_\b-\frac{i}{2}(n_\a-n_\b)|^2}
     {|q_\a-q_\b-\frac{i}{2}(n_\a+n_\b)|^2}
\prod_{\a=1}^M\frac{e^{in_{\a}q_{\a}x}}{n_\a}
\notag\\
&~~\times\prod_{r=1}^{n_\a}
\frac{1}{(-iq_{\a}+v_-+\frac12(n_{\a}-2r))(iq_\a+v_++\frac{1}{2}(n_\a-2r))}.
\label{integrand1}
\end{align}

We can further deform~\eqref{integrand1} to an
expression in terms of a determinant by 
using the Cauchy's determinant formula,
\begin{equation}
\frac{\prod_{\a<\b}^M(a_\a-a_\b)(b_\a-b_\b)}
{\prod_{\a,\b=1}^M(a_{\a}-b_\b)}=(-1)^{\frac{M(M-1)}{2}}
\det\left(\frac{1}{a_\a-b_\b}\right),
\end{equation}
and a few basic properties of determinant. We find
\begin{align}
\langle x|\Psi_z\rangle\langle\Psi_z|\Phi\rangle
&=2^MN!\prod_{\ell=1}^N(v_++v_--\ell)
\det\left(\frac{1}{n_j+n_k+2i(q_j-q_k)}\right)_{j,k=1}^M\notag\\
&~~\times\prod_{\a=1}^Me^{in_{\a}q_\a x}\prod_{r=1}^{n_\a}
\frac{1}{(-iq_{\a}+v_-+\frac12(n_{\a}-2r))(iq_\a+v_++\frac{1}{2}(n_\a-2r))}
\notag\\
&=2^MN!\prod_{\ell=1}^N(v_++v_--\ell)
\prod_{\a=1}^M\left(\int_0^{\infty}d\o_\a\right)
\det\left(e^{-\o_j(n_j+n_k+2i(q_j-q_k))}\right)_{j,k=1}^M\notag\\
&~~\times\prod_{\a=1}^Me^{in_{\a}q_\a x}\prod_{r=1}^{n_\a}
\frac{1}{(-iq_{\a}+v_-+\frac12(n_{\a}-2r))(iq_\a+v_++\frac{1}{2}(n_\a-2r))}
\notag\\
&=2^MN!\prod_{\ell=1}^N(v_++v_--\ell)\prod_{\a=1}^M \left(\int_0^{\infty}d\o_\a\right)
\det\left(K_{x}(n_j,q_j;\o_j,\o_k)
\right)_{j,k=1}^M,
\label{integrand2}
\end{align}
where 
in the last equality, we used a simple fact
\begin{equation}
 \det(a_j^{b_j+b_k}) = \det((a_j a_k)^{b_j}).
\end{equation}
and the notation
\begin{align}
&~K_{x}(n_j,q_j;\o_j,\o_k)\notag\\
&:=e^{in_jq_jx-n_j(\o_j+\o_k)-2iq_j(\o_j-\o_k)}
\prod_{r=1}^{n_j}
\frac{1}{(-iq_j+v_-+\frac12(n_j-2r))(iq_j+v_++\frac{1}{2}(n_j-2r))}
\notag\\
&=e^{in_jq_jx-n_j(\o_j+\o_k)-2iq_j(\o_j-\o_k)}
\frac{\Gamma\left(-iq_j+v_--\frac{n_j}{2}\right)
        \Gamma\left(iq_j+v_+-\frac{n_j}{2}\right) }
       {\Gamma\left(-iq_j+v_-+\frac{n_j}{2}\right)
        \Gamma\left(iq+v_++\frac{n_j}{2}\right)}.
\label{Kxdef}
\end{align}

From~\eqref{eigenvalue},~\eqref{afterdeform}, 
and~\eqref{integrand2},
we obtain an expression of $\langle Z^N(x,t)\rangle$ in terms of
the determinant,
\begin{align}
&\quad\langle Z^N(x,t)\rangle e^{\frac{Nt}{24}+\frac{Nx^2}{2t}}
\notag\\
&=\sum_{M=1}^N\frac{2^MN!}{M!}\prod_{\ell=1}^N(v_++v_--\ell)\prod_{\a=1}^M
\left(\sum_{n_\a=1}^{\infty}
\int_{-\infty}^{\infty}\frac{dq_\a}{2\pi}
e^{-\frac{t}{2}n_\a q_\a^2+\frac{t}{24}n_\a^3}
\int_0^\infty d\o_\a\right)
\delta_{\sum_{\beta=1}^Mn_\beta,N}\notag\\
&~\times
\det\left(e^{n_j\frac{x^2}{2t}}K_{x}(n_j,q_j;\o_j,\o_k)
\right)_{j,k=1}^M\notag\\
&=\sum_{M=1}^N\frac{N!}{M!}\prod_{\ell=1}^N(v_++v_--\ell)\prod_{\a=1}^M\left(
\int_0^\infty d\o_\a\sum_{n_\a=1}^{\infty}\right)\delta_{\sum_{\beta=1}^M
n_\beta,N}\notag\\
&\hspace{4cm}\times\det\left(\int_{-\infty}^{\infty}\frac{dq}{\pi}
e^{n_j\frac{x^2}{2t}-\frac{t}{2}n_jq^2+\frac{t}{24}n_j^3}K_{x}(n_j,q;\o_j,\o_k)
\right)_{j,k=1}^M.
\label{zndet}
\end{align}
Here we would like to discuss the condition on the drifts $v_{\pm}$.
In the above equation, we notice that the condition 
$v_{\pm}>N/2+\max_{\a}n_\a/2$ discussed below~\eqref{nodangerous}
ensures that all the poles 
$q=-i(v_-+(n_j/2-r)$ and $q=i(v_+-(n_j/2-r)$ with $~r=0,1,\cdots,n_j$ 
in the integrand $e^{n_j\frac{x^2}{2t}-\frac{t}{2}n_jq^2+\frac{t}{24}n_j^3}K_{x}(n_j,q;\o_j,\o_k)$ lie on the lower and upper half plane respectively. 
Thus we can relax the condition to $v_{\pm}>\max_{\a}n_\a/2$, where
the above property of the poles is retained.
\section{Generating function and height distribution}\label{gfhd}
\subsection{Generating function}
Now we define the generating function $G_{\gamma_t}(s;X)$ as 
\begin{equation}
G_{\gamma_t}(s;X)
=
\sum_{N=0}^{\infty}\frac{\left(-e^{-\gamma_t s}\right)^N}{N!}
\left\langle 
{Z}^N \left( 2\gamma_t^2X,t \right)
\right\rangle  
e^{N\frac{\gamma_t^3}{12} +N\gamma_t X^2} 
%\tilde{Z}_{\nu,\lambda,D}^N(2\gamma_t^2X,t)\rangle
=
\langle e^{-e^{\gamma_t(H_t(X)-s)}}\rangle,
\label{grm}
\end{equation}
where in the last equality, we used~\eqref{scaled}. 
Substituting~\eqref{zndet} into this equation,
we get 
\begin{align}
&~G_{\gamma_t}(s;X) %:=\sum_{N=0}^\infty\frac{(-e^{-\gamma_t s})^N}
\notag\\
&=1+\sum_{N=1}^\infty\prod_{\ell=1}^N(v_++v_--\ell)\sum_{M=1}^N\frac{(-e^{-\gamma_t s})^N}{M!}
\prod_{\a=1}^M\left(
\int_0^\infty d\o_\a
\sum_{n_\a=1}^{\infty}\right)\delta_{\sum_{\beta=1}^Mn_\beta,N}
\notag\\
&~\times\det\left(
\int_{-\infty}^{\infty}\frac{dq}{\pi}
e^{n_j\gamma_tX^2-\gamma_t^3n_jq^2+\frac{\gamma_t^3}{12}n_j^3}
K_{2\gamma_t^2X}(n_j,q;\o_j,\o_k)
\right)_{j,k=1}^M
\label{KPZfred2}.
\end{align}

\subsection{Shifting}\label{Shifting}
In the previous studies for the narrow-wedge~\cite{D2010,CLDR2010} and 
half Brownian motion initial condition~\cite{IS2011}, the generating functions 
corresponding to~\eqref{KPZfred2}
were already described by the Fredholm determinant, 
which is an important achievement
in the replica analyses. 
Unfortunately, however, this is not the case in~\eqref{KPZfred2}
due to the factor $\prod_{j=1}^N(v_++v_--j)$. 
In order to resolve this problem, 
we introduced $\tilde{h}(x,t)$
in~\eqref{tildechi}. Here we discuss some general properties of
two random variables $h$ and $\tilde{h}$ related as
\begin{equation}
 \tilde{h} = h + \chi, 
\end{equation}
where $\chi$ is also a random variable independent of $h$.  Note 
that at this stage, $h$ and $\tilde{h}$ are just two random variables
and do not have a specific meaning such as height.  
In addition, right now we do not assume a specific distribution 
for $\chi$. Let us define
\begin{align}
Z=e^{h},~M(\xi)=\sum_{N=0}^{\infty}\frac{\langle (\xi h)^N\rangle}{N!}
=\left\langle e^{\xi h}\right\rangle,
~F(s)=\P [h\le s].
\label{ZGF}
\end{align}
and also
$\tilde{Z},~\tilde{M}(\xi),~\tilde{F}(s)$ in the same way as~\eqref{ZGF} except
that $h$ is replaced by $\tilde{h}$. We find that the following relation
holds
\begin{align}
&\langle Z^N\rangle=\frac{1}{g(N)}\langle \tilde{Z}^N\rangle
\label{ztz}
\\
&M(\xi)=\frac{1}{g(\xi)}\tilde{M}(\xi),
\end{align}
and noticing the fact
$
\frac{d F(s)}{ds}=\int_{-\infty}^{\infty}d\xi M(i\xi)e^{-i\xi s},
$
we also see
\begin{align}
F(s)=\frac{1}{g(-\partial/\partial s)}\tilde{F}(s)
\label{tFFr}
\end{align}
where $g(y)=\langle e^{y\chi}\rangle$.

In~\eqref{tildechi}, we chose $\chi$ as the random variable 
such that $e^{\chi}$ obey the inverse gamma distribution with parameter 
$v_++v_-$.
The probability density function of the inverse gamma random variable 
is 
\begin{equation}
 p_\theta(x) = \frac{1}{\Gamma(\theta)}x^{-\theta-1}e^{-1/x} {\bf 1}_{x\geq 0},
\end{equation}
where $\theta >0$ is a parameter and the function $g(\xi)$ in this case 
is given by
\begin{equation}
 g(\xi)=\int_0^{\infty} x^{\xi} p_{\theta}(x)dx
 =
 \frac{1}{\Gamma(\theta)}\int_0^{\infty} x^{\xi-\theta-1} e^{-1/x} dx
 =
 \frac{\Gamma(\theta-\xi)}{\Gamma(\theta)}.
 %=
 %\frac{1}{(\theta-1)(\theta-2)\cdots (\theta-n)}.
\label{iglt}
\end{equation}
In particular, we notice that if we set $\theta=v_++v_-$ as in~\eqref{scaled}, 
$g(N)$ becomes $1/\prod_{j=1}^N (v_++v_--j)$.

\subsection{Generating function for the generalized height}
We define $\tilde{Z}(x,t)$ and $\tilde{G}_{\gamma_t}(s;X)$ 
as~\eqref{zdef} and~\eqref{grm} respectively with $h(x,t)$ replaced by
$\tilde{h}(x,t)$. From the relation~\eqref{ztz} and~\eqref{iglt} with 
$\theta=v_++v_-$, we see that $\langle\tilde{Z}^N(x,t)\rangle$ 
is related to $\langle Z^N(x,t)\rangle$ as
\begin{align}
\langle\tilde{Z}^N(x,t)\rangle=\langle Z^N(x,t)\rangle\frac{1}{\prod_{j=1}^N (v_++v_--j)}.
\end{align}
Hence $\tilde{G}_{\gamma_t}(s;X)$ is calculated as
\begin{align}
&~\tilde{G}_{\gamma_t}(s;X) :=\sum_{N=0}^\infty\frac{(-e^{-\gamma_t s})^N}
{N!}\langle\tilde{Z}^N(2\gamma_t^2X,t)\rangle
\notag\\
&=1+\sum_{N=1}^\infty\sum_{M=1}^N\frac{(-e^{-\gamma_t s})^N}{M!}
\prod_{\a=1}^M\left(
\int_0^\infty d\o_\a
\sum_{n_\a=1}^{\infty}\right)\delta_{\sum_{\beta=1}^Mn_\beta,N}\notag\\
&\hspace{5cm}\times\det\left(
\int_{-\infty}^{\infty}\frac{dq}{\pi}
e^{n_j\gamma_tX^2-\gamma_t^3n_jq^2+\frac{\gamma_t^3}{12}n_j^3}
K_{2\gamma_t^2X}(n_j,q;\o_j,\o_k)
\right)_{j,k=1}^M
\notag\\
\end{align}
Note that this generating function does not include the term
$\prod_{\ell=1}^N(v_++v_--\ell)$ in~\eqref{KPZfred2}. 
In this case, we find the Fredholm determinant expression, 
\begin{align}
&~\tilde{G}_{\gamma_t}(s;X)
=\sum_{M=0}^\infty\frac{(-1)^M}{M!}
\prod_{\a=1}^M
\int_0^\infty d\o_\a\notag\\
&\hspace{1cm}\times\det\left(
\sum_{n=1}^{\infty}(-1)^{n-1}
\int_{-\infty}^{\infty}\frac{dq}{\pi}
e^{n_j\gamma_tX^2-\gamma_t^3n_jq^2+\frac{\gamma_t^3}{12}n_j^3-\gamma_tns}K_{2\gamma_t^2X}
(n,q;\o_j,\o_k)
\right)_{j,k=1}^M
\label{KPZfred2t}.
\end{align}
Shifting the variable $q$ to $q+iX/\gamma_t$, we get
\begin{align}
&~\tilde{G}_{\gamma_t}(s;X)
=\sum_{M=0}^\infty\frac{(-1)^M}{M!}
\prod_{k=1}^M
\int_0^\infty d\o_k
\det\left(
K_{v_+-\frac{X}{\gamma_t},v_-+\frac{X}{\gamma_t}}(\o_j,\o_k)
\right)_{j,k=1}^M
\label{KPZshift},
\end{align}
where the kernel $K_{v,w}(\o_j,\o_k)$ is written as
\begin{align}
K_{v,w}(\o_j,\o_k)&=
\sum_{n=1}^{\infty}(-1)^{n-1}
\int_{-\infty}^{\infty}\frac{dq}{\pi}
e^{-n(\o_j+\o_k)-2iq(\o_j-\o_k)-\gamma_t^3nq^2+\frac{\gamma_t^3}{12}n^3
-\gamma_t ns}\notag\\
&\hspace{4cm}\times
\frac{\Gamma\left(-iq+w-\frac{n}{2}\right)
        \Gamma\left(iq+v-\frac{n}{2}\right) }
       {\Gamma\left(-iq+w+\frac{n}{2}\right)
        \Gamma\left(iq+v+\frac{n}{2}\right)},
\label{kernelkvw}
\end{align}
with $v,w>n/2$. In~\eqref{KPZshift}, we keep
the contour of $q$ on the real axis by changing 
the condition for the drift to 
$v_+-X/\gamma_t>n/2,~v_-+X/\gamma_t>n/2$.

In the following, we provide another representation with 
the kernel using the deformed Airy functions
$\Ai_{\Gamma}^{\Gamma}(a,b,c,d)$ defined in~\eqref{AiGGm}.
We use the following two relations,

\vspace{5mm}
\noindent
{\bf Lemma 6}

\noindent
(a)
{\it 
We set $q\in\R$ and ~$m,~n\ge0$. 
When $a,b > n/2$, we have
\begin{align}
\frac
{\Gamma\left(-iq+a-\frac{n}{2}\right)\Gamma\left(iq+b-\frac{n}{2}\right)}
{\Gamma\left(-iq+a+\frac{n}{2}\right)\Gamma\left(iq+b+\frac{n}{2}\right)} 
e^{\frac{m^3n^3}{3}}
=\int_{-\infty}^{\infty}dy
\Ai_{\Gamma\Gamma}^{\Gamma\Gamma}
\left(y,\frac{1}{2m},-iq+a,iq+b\right) e^{mny},
\label{lemma6(1)}
\end{align}
where
\begin{align}
\Ai_{\Gamma\Gamma}^{\Gamma\Gamma}\left(a,b,c,d\right)
=
\frac{1}{2\pi}
\int_{\Gamma_{i\frac{c,d}{b}}}dz e^{iaz+iz^3/3}
\frac{\Gamma(ibz+c)\Gamma(ibz+d)}{\Gamma(-ibz+c)\Gamma(-ibz+d)}
\label{AiGG}
\end{align}
with $\Gamma_p$ defined in~\eqref{AiG2} below.

\noindent
{\rm (b)} For $a,b,x\in\R$ and $w\ge 0$, we have
\begin{align}
&~\frac{1}{2\pi}\int_{-\infty}^{\infty}dp
\Ai_{\Gamma\Gamma}^{\Gamma\Gamma}
\left(p^2+v,w,-iwp+a,iwp+b\right)e^{ipx}\notag\\
&=
\frac{1}{2^{\frac{1}{3}}}
\Ai_\Gamma^\Gamma\left(2^{-\frac{2}{3}}(v+x),2^{\frac{1}{3}}w,a,b\right)
\Ai_\Gamma^\Gamma\left(2^{-\frac{2}{3}}(v-x),2^{\frac{1}{3}}w,b,a\right),
\label{lemma6(2)}
\end{align}
where $\Ai^\Gamma(a,b,c)$ and $\Ai_\Gamma(a,b,c)$ are defined by~\eqref{AiG1} 
and~\eqref{AiG2} respectively.
}

\vspace{3mm}
As will be discussed in~Sec.~\ref{Co}, these relations~\eqref{lemma6(1)} 
and~\eqref{lemma6(2)} play a crucial role in the replica analysis.
These are two-parameters ($a,b$) generalization 
of the ones used in the previous studies of half-Brownian motion~\cite{IS2011} and the narrow wedge~\cite{CLDR2010,D2010}
initial data: As either $a$ or $b$ goes to infinity in~\eqref{lemma6(1)}
and~\eqref{lemma6(2)}, they reproduce Lemma~6 (a) and (b) 
in~\cite{IS2011} respectively. On the other hand, in the case where both $a$ and $b$ go to infinity, they converge to the ones appeared in~\cite{CLDR2010,D2010}.

\vspace{3mm}
\noindent
{\bf Proof}

\noindent
(a)
RHS of~\eqref{lemma6(1)} is written as
\begin{align}
\frac{1}{2\pi}\int_{-\infty}^{\infty}dy\int_{\R+imn}dz
\frac{\Gamma\left(i\frac{z}{2m}-iq+a\right)
\Gamma\left(i\frac{z}{2m}+iq+b\right)}
{\Gamma\left(-i\frac{z}{2m}-iq+a\right)
\Gamma\left(-i\frac{z}{2m}+iq+b\right)}
e^{i(z-imn)y+i\frac{z^3}{3}}.
\end{align}
Here we used the fact that 
the contour of $z$ can be deformed from to 
$\R+imn$ since the imaginary parts of the poles 
of the integrand $2m(ia-q+ir)$ and $2m(ib-q+ir)$, where 
$r=0,1,2,\cdots$ are larger than $mn$ when the condition $a,b>n/2$ 
is satisfied. In this
equation, we change the variable $z$ on RHS to $y_2=z-imn$
and get
\begin{align}
&~\frac{1}{2\pi}
\int_{-\infty}^{\infty}dy\int_{-\infty}^{\infty}d{y}_2 
\frac{\Gamma\left(-iq+a-\frac{n}{2}+i\frac{{y}_2}{2m}\right)
\Gamma\left(iq+b-\frac{n}{2}+i\frac{{y}_2}{2m}\right)
}
{\Gamma\left(-iq+a+\frac{n}{2}-i\frac{{y}_2}{2m}\right)
\Gamma\left(iq+b+\frac{n}{2}-i\frac{{y}_2}{2m}\right)}
e^{iy{y}_2+i\frac{({y}_2+im n)^3}{3}}\notag\\
&=\int_{-\infty}^{\infty}d{y}_2\delta({y}_2)
\frac{\Gamma\left(-iq+a-\frac{n}{2}+i\frac{{y}_2}{2m}\right)
\Gamma\left(iq+b-\frac{n}{2}+i\frac{{y}_2}{2m}\right)
}
{\Gamma\left(-iq+a+\frac{n}{2}-i\frac{{y}_2}{2m}\right)
\Gamma\left(iq+b+\frac{n}{2}-i\frac{{y}_2}{2m}\right)}
e^{i\frac{({y}_2+im n)^3}{3}}\notag\\
&=\frac{\Gamma\left(-iq+a-\frac{n}{2}\right)\Gamma\left(iq+b-\frac{n}{2}\right)}
{\Gamma\left(-iq+a+\frac{n}{2}\right)\Gamma\left(iq+b+\frac{n}{2}\right)}
e^{\frac{m^3 n^3}{3}}.
\end{align}
\noindent
(b)
LHS of~\eqref{lemma6(2)} reads
\begin{align}
\frac{1}{(2\pi)^2}\int_{-\infty}^{\infty} dp
\int_{\Gamma_{-p+i\frac{a,b}{w}}}dz e^{i(p^2+v)z+i\frac{z^3}{3}+ipx}
\frac{\Gamma\left(iw(z-p)+a\right) \Gamma\left(iw(z+p)+b\right) }
{\Gamma\left(-iw(z+p)+a\right)\Gamma\left(-iw(z-p)+b\right)}.
\end{align}
By applying the change of variables $p=(z_1-z_2)/2^{2/3}$
and $z=(z_1+z_2)/2^{2/3}$, we obtain the desired expression.
\qed

\vspace{5mm}
Applying (a) of the lemma with $a=v_-+X/\gamma_t,b=v_+-X/\gamma_t$ 
and $m=\gamma_t/2^{2/3}$ 
to~\eqref{kernelkvw}, one has
\begin{align}
&~~K_{v_+-\frac{X}{\gamma_t},v_-+\frac{X}{\gamma_t}}(\o_j,\o_k)\notag\\
&=\sum_{n=1}^{\infty}(-1)^{n-1}
\int_{-\infty}^{\infty}\frac{dq}{\pi}
\int_{-\infty}^{\infty}dy_1
\Ai_{\Gamma\Gamma}^{\Gamma\Gamma}
\left(y_1,\frac{1}{2^{\frac{1}{3}}\gamma_t},
      -iq+v_-+\frac{X}{\gamma_t}, iq+v_+-\frac{X}{\gamma_t}
\right)
e^{-2iq(\o_j-\o_k)}\notag\\
&
\hspace{5cm}\times e^{-n(\o_j+\o_k+\gamma_t^3q^2+\gamma_ts-2^{-\frac{2}{3}}\gamma_ty_1)}.
\label{aiggggex}
\end{align}
So far we have imposed  the condition for the drifts 
$v_{+}-X, v_{-}+X>n/2$ (see~\eqref{KPZshift} and~\eqref{kernelkvw}).
At this point, since RHS of~\eqref{lemma6(1)} is well-defined for 
all $a,b\in\R$, we can relax it to the region such that the generalized
height~\eqref{tildechi} under consideration is well-defined, i.e. $v_++v_->0$.
Changing the variables $q,~\o_j$ and $y_1$ to 
$p=2^{1/3}\gamma_t q,~\xi_j=2\gamma_t^{-1}\o_j$ and
$y=y_1/2^{2/3}-p^2/2^{2/3}-(\xi_j+\xi_k)/2$,
we see 
$K_{v_+-X/\gamma_t,v_-+X/\gamma_t}(\o_j,\o_k)d\o_j
=\bar{K}_{v_+-X/\gamma_t,v_-+X/\gamma_t}(\xi_j,\xi_k)d\xi_j$, where
\begin{align}
\bar{K}_{v,w}(\xi_j,\xi_k)
&=\frac{2^{\frac{1}{3}}}{2\pi}
\int_{-\infty}^{\infty}\hspace{-1mm}dy
\int_{-\infty}^{\infty}\hspace{-1mm}{dp}
\Ai_{\Gamma\Gamma}^{\Gamma\Gamma}
\left(p^2+2^{\frac{2}{3}}y+\frac{\xi_j+\xi_k}{2^{\frac{1}{3}}},
\frac{1}{2^{\frac{1}{3}}\gamma_t},
\frac{-ip+2^{\frac13} \gamma_t w }{2^{\frac{1}{3}}\gamma_t},
\frac{ip+2^{\frac13} \gamma_t v }{2^{\frac{1}{3}}\gamma_t}
\right)\notag\\
&\hspace{4cm}\times e^{-i\frac{(\xi_j-\xi_k)}{2^{\frac{1}{3}}}p}
%\left(,\frac{1}{2^{1/3}\gamma_t},iq-\frac{X}{\gamma_t}
%\right)
\sum_{n=1}^{\infty}(-1)^{n-1}e^{-\gamma_tn(s-y)}
\notag\\
&=
\int_{-\infty}^{\infty} \hspace{-1mm}dy 
\Ai_{\Gamma}^\Gamma
\left(\xi_j+y,\frac{1}{\gamma_t},
      v,w\right)
\Ai_{\Gamma}^{\Gamma}
\left(\xi_k+y,\frac{1}{\gamma_t},
      w,v \right)
\frac{e^{\gamma_t y}}{e^{\gamma_t y}+e^{\gamma_t s}}.
\label{alternative1m}
\end{align}
where in the second equality we applied (2) of Lemma~6 to this equation 
with $v=2^{2/3}y+(\xi_j+\xi_k)/2^{1/3}$, $w=1/(2^{1/3}\gamma_t)$, 
$a=v_-+X/\gamma_t,~b=v_+-X/\gamma_t$, $x=-(\xi_j-\xi_k)/2^{1/3}$.
Note that although the sum 
$\sum_{n=1}^{\infty}(-1)^{n-1}e^{-\gamma_tn(s-y)}$
in the first equality become divergent for $s-y<0$,
the last expression 
$e^{\gamma_ty}/(e^{\gamma_ty}+e^{\gamma_ts})$
in the second equality is well-defined in arbitrary
$s-y\in\R$. As we will comment in Sec.~\ref{Co}, this divergence originates
from the ill-defined nature of the generating function and
this kind of analytic continuation has been necessary 
for the previous studies on the replica analyses~\cite{CLDR2010,D2010}.
Using~\eqref{alternative1m}, we eventually obtain
\begin{align}
\tilde{G}_{\gamma_t}(s;X)=\det\left(1-P_0\bar{K}_{v_+-\frac{X}{\gamma_t},
v_-+\frac{X}{\gamma_t}}P_0
\right).
\label{KPZfred1m}
\end{align}

\subsection{Proofs of Proposition 1}
Eq.~\eqref{Result} follows immediately from~\eqref{tFFr} and~\eqref{iglt}.
For the derivation of~\eqref{Pr1}, we use 
a formula given in~\cite{CLDR2010,PS2010p,IS2011} which transform
the generating function $\tilde{G}_{\gamma_t}(s;X)$ into 
the height distribution $\tilde{F}_{v_+,v_-,t}(s;X)$~\eqref{hdist}.
By using the Fredholm determinant~\eqref{KPZfred1m}, 
$\tilde{F}_{\gamma_t}(s;X)$ can be expressed as 
follows~\cite{CLDR2010,PS2010p}.
\begin{align}
\tilde{F}_{v_{\pm},t}(s;X)
=1-\int_{-\infty}^{\infty}
du e^{-e^{\gamma_t(s-u)}}
g_{\gamma_t}(u;X).
\end{align}
Here
\begin{align}
g_{\gamma_t}(u;X)=\frac{1}{2\pi i}
\left(
\det(1-P_0K^+_{v_+-\frac{X}{\gamma_t},v_-+\frac{X}{\gamma_t}}P_0)
-\det(1-P_0K^-_{v_+-\frac{X}{\gamma_t},v_-+\frac{X}{\gamma_t}}P_0)
\right)
\label{gexpressionm}
\end{align}
where $K^{\pm}_{v_+-\frac{X}{\gamma_t},v_-+\frac{X}{\gamma_t}}(x,y)$ 
is the kernel~\eqref{alternative1m}
in which the term $e^{-s}$ is replaced by $-e^{u}\pm i\e$
with $\e>0$ being infinitesimal.

Using~\eqref{alternative1m}, and the relation
$1/(x\pm i\e)= \mathcal{P}(1/x)\mp i\pi\delta(x)$, where
$\mathcal{P}$ denotes the Cauchy principal value,
we can easily find
$K^{\pm}_X$ in~\eqref{gexpressionm} is represented as
\begin{align}
K^{\pm}_X(\xi_j,\xi_k)
&=\mathcal{P}\int_{-\infty}^{\infty} dy
\frac{1}{1-e^{\gamma_t(u-y)}}\notag\\ 
&\hspace{1cm}\times\Ai_{\Gamma}^\Gamma\left(\xi_j+y,\frac{1}{\gamma_t},
     v_+-\frac{X}{\gamma_t},v_-+\frac{X}{\gamma_t}\right)
\Ai_{\Gamma}^{\Gamma} \left(\xi_k+y,\frac{1}{\gamma_t},
     v_-+\frac{X}{\gamma_t},v_+-\frac{X}{\gamma_t}\right)
\notag\\
&~~~~
\mp i\pi 
\Ai_{\Gamma}^\Gamma\left(\xi_j+u,\frac{1}{\gamma_t},
     v_+-\frac{X}{\gamma_t},v_-+\frac{X}{\gamma_t}\right)
\Ai_{\Gamma}^{\Gamma} \left(\xi_k+u, \frac{1}{\gamma_t},
    v_-+\frac{X}{\gamma_t},v_+-\frac{X}{\gamma_t}\right).
\end{align}
Substituting this expression to~\eqref{gexpressionm} and using
basic properties of determinant, we eventually arrive
at the expression~\eqref{Th3g} in Proposition 1.

\section{Stationary limit}\label{sl}
In this section, we provide the proof of Theorem 2, which represents
the exact height distribution in the stationary limit~\eqref{defFw}. 
This is obtained by rewriting the representation in Proposition 1
in a more suitable form for taking this limit. 

We first focus on the part
$\det\left(1-P_u(B^{\Gamma}_{\gamma_t}-P^{\Gamma}_{\Ai})P_u\right)$
in~\eqref{Th3g}. 
Introducing the scaled drifts $\o_{\pm}$ as 
$v_{\pm}=\o_{\pm}/\gamma_t$,
one finds that it is written as 
\begin{align}
&B^{\Gamma}_{\gamma_t}(\xi_1,\xi_2)-P^{\Gamma}_{\Ai}(\xi_1,\xi_2)\notag\\
&=
\int_{-\infty}^{\infty}dy
C_t^{(\delta)}(y)
\Ai_\Gamma^\Gamma\left(\xi_1+y,
\frac{1}{\gamma_t},\frac{\o_+-X}{\gamma_t},\frac{\o_-+X}{\gamma_t}\right)
\Ai_\Gamma^\Gamma \left(\xi_2+y,
\frac{1}{\gamma_t},\frac{\o_-+X}{\gamma_t},\frac{\o_+-X}{\gamma_t}\right),
\end{align}
where $C_t^{(\delta)}(y)$ is defined after~\eqref{kerneld}. 
Thus it is expressed
by a product of integral kernels 
$P_u(B^{\Gamma}_{\gamma_t}-P^{\Gamma}_{\Ai})P_u(\xi_1,\xi_2)=
X_1X_2(\xi_1,\xi_2)$
where 
\begin{align}
&X_1(\xi_1,y)=P_u(\xi_1)\Ai_\Gamma^\Gamma\left(\xi_1+y,
\frac{1}{\gamma_t},\frac{\o_+-X}{\gamma_t},\frac{\o_-+X}{\gamma_t}\right),\\
&X_2(y,\xi_2)=C_t^{(\delta)}(y)\Ai_\Gamma^\Gamma \left(\xi_2+y,
\frac{1}{\gamma_t},\frac{\o_-+X}{\gamma_t},\frac{\o_+-X}{\gamma_t}\right)
P_u(\xi_2)
\end{align}
By using the relation of the Fredholm determinant 
$\det(1-X_1X_2)=\det(1-X_2X_1)$, we have 
\begin{align}
\det\left(1-P_u(B^{\Gamma}_{\gamma_t}-P^{\Gamma}_{\Ai})P_u\right)=
\det(1-X_2X_1)
\end{align}
where the kernel $X_2X_1$ is expressed as
\begin{align}
&~X_2X_1(y_1,y_2)\notag\\
&=C_t^{(\delta)}(y_1)\int_{u}^{\infty}d\xi
\Ai_\Gamma^\Gamma \left(y_1+\xi,
\frac{1}{\gamma_t},\frac{\o_-+X}{\gamma_t},\frac{\o_+-X}{\gamma_t}\right)
\Ai_\Gamma^\Gamma\left(y_2+\xi,
\frac{1}{\gamma_t},\frac{\o_+-X}{\gamma_t},\frac{\o_-+X}{\gamma_t}\right)
\notag\\
&=\frac{C_t^{(\delta)}(y_1)}{(2\pi)^2}\int_{\Gamma_{i(\o_+-X)}} 
\hspace{-8mm}dx_1
\int_{\Gamma_{i(\o_-+X)}}  
\hspace{-8mm}
dx_2 e^{ix_1(y_1+u)+ix_2(y_2+u)+\frac{i}{3}(x_1^3+x_2^3)}
\frac{\Gamma\left(\frac{\o_+-X+ix_1}{\gamma_t}+1\right)
\Gamma\left(\frac{\o_-+X+ix_2}{\gamma_t}+1\right)}
{\Gamma\left(\frac{\o_-+X-ix_1}{\gamma_t}+1\right)
\Gamma\left(\frac{\o_+-X-ix_2}{\gamma_t}+1\right)}\notag\\
&\hspace{3cm}\times\frac{-1}{i(x_1+x_2)}\frac{\o_-+X-ix_1}{\o_+-X+ix_1}
\frac{\o_+-X-ix_2}{\o_-+X+ix_2}.
\end{align}
Noticing the relation 
\begin{align}
\frac{1}{i(x_1+x_2)}\frac{b-ix_1}{a+ix_1}
\frac{a-ix_2}{b+ix_2}
=\frac{1}{i(x_1+x_2)}-(a+b)\frac{1}{ix_1+a}\frac{1}{ix_2+b}
\end{align}
we can further rewrite it as
\begin{align}
&X_2X_1(y_1,y_2)\notag\\
&=
C_t^{(\delta)}(y_1)\int_{u}^{\infty}d\xi
\Ai_\Gamma^\Gamma \left(y_1+\xi,
\frac{1}{\gamma_t},1+\frac{\o_-+X}{\gamma_t},1+\frac{\o_+-X}{\gamma_t}\right)
\notag\\
&\hspace{3cm}\times\Ai_\Gamma^\Gamma\left(y_2+\xi,
\frac{1}{\gamma_t},1+\frac{\o_+-X}{\gamma_t},1+\frac{\o_-+X}{\gamma_t}\right)
\notag\\
&+(\o_++\o_-)C_t^{(\delta)}(y_1)B_{\o_+-X,\o_-+X,u}(y_1)B_{\o_-+X,\o_+-X,u}(y_2),
\end{align}
where $B_{a,b,u}(x)$ is represented as
\begin{align}
B_{a,b,u}(x)=\frac{-1}{2\pi}\int_{\Gamma_{ia}} dz
\frac{e^{iz(x+u)+\frac{i}{3}z^3}}{iz+a}
\frac{\Gamma\left(\frac{a+iz}{\gamma_t}+1\right)}
{\Gamma\left(\frac{b-iz}{\gamma_t}+1\right)}
\end{align}
and we find this can be rewritten as~\eqref{defB}. 

Thus we have
\begin{align}
&\det\left(1-P_u(B^{\Gamma}_{\gamma_t}-P^{\Gamma}_{\Ai})P_u\right)\notag\\
&=\det\left(1-A^{(\delta)}_{\o_+-X,\o_-+X}\right)
\left(1-(\o_++\o_-)\int_{-\infty}^{\infty}dx
(\rho_{A^{(\delta)}}C_t^{(\delta)}B_{\o_+-X,\o_-+X,u})(x)B_{\o_-+X,\o_+-X,u}(x)\right)\notag\\
&=(\o_++\o_-)\det\left(1-A^{(\delta)}_{\o_+-X,\o_-+X}\right)\notag\\
&~~\times\left(\frac{1-\o_+-\o_-}{\o_++\o_-}-\int_{-\infty}^{\infty}dx
(C_t^{(\delta)}B_{\o_+-X,\o_-+X,u})(x)B_{\o_-+X,\o_+-X,u}(x)\right.\notag\\
&\hspace{1cm}\left.+1-\int_{-\infty}^{\infty}dx (A^{(\delta)}
\rho_{A^{(\delta)}}C_t^{(\delta)}B_{\o_+-X,\o_-+X,u})(x)B_{\o_-+X,\o_+-X,u}(x)
\right)\notag\\
&=(\o_++\o_-)\left(\det\left(1-A^{(\delta)}_{\o_+-X,\o_-+X}\right)
L^{(\delta)}_{\o_+-X,\o_-+X}(u)-\det(1-A^{(\delta)}_{\o_+-X,\o_-+X}
-D^{(\delta)}_{\o_+-X,\o_-+X})
\right).
\label{stlimit1}
\end{align}
Here $A^{(\delta)}_{a,b}$, 
$D^{(\delta)}_{a,b}$ and $L^{(\delta)}_{a,b}$ are defined in
Theorem 3 and $\rho_{A^{(\delta)}}(x,y):=(1-A^{(\delta)})^{-1}(x,y)$.
We can also rewrite the other term, 
$\det\left(1-P_uB^{\Gamma}_{\gamma_t}P_u\right)$ in~\eqref{Th3g} 
in the completely 
parallel way. We have
\begin{align}
&~\det\left(1-P_uB^{\Gamma}_{\gamma_t}P_u\right)\notag\\
&=(\o_++\o_-)\left(\det\left(1-A_{\o_+-X,\o_-+X}\right)
L_{\o_+-X,\o_-+X}(u)-\det(1-A_{\o_+-X,\o_-+X}
-D_{\o_+-X,\o_-+X})\right)
\label{stlimit2}
\end{align}
Using these forms~\eqref{stlimit1} and~\eqref{stlimit2},
we finally obtain the desired expression~\eqref{Result2},
\begin{align}
&F_{w,t}(s,X)=\lim_{\substack{\o_-\rightarrow -\o_+\\ \o_+=w}}
F_{\frac{\o_+}{\gamma_t},\frac{\o_-}{\gamma_t},t}(s;X)\notag\\
%:={\rm{Prob}}\left(
%h(x,t)+\gamma_t^3/12+\gamma_tX^2\le \gamma_t s \right)
&=\lim_{
\substack{\o_-\rightarrow -\o_+\\ \o_+=w}
}
\frac{\Gamma\left(\frac{\o_++\o_-}{\gamma_t}+1\right)}
{\Gamma\left(\frac{\o_++\o_-+d/ds}{\gamma_t}+1\right)}
\left(1+\frac{d/ds}{\o_++\o_-}\right)
\tilde{F}_{\frac{\o_+}{\gamma_t},\frac{\o_-}{\gamma_t},t}(s;X)
%\left(1-\int_{-\infty}^{\infty}
%du e^{-e^{\gamma_t(s-u)}}
%g_{\gamma_t}(u;X)\right)
\notag\\
&=
\lim_{
\substack{\o_-\rightarrow -\o_+\\ \o_+=w}
}
\frac{\Gamma\left(\frac{\o_++\o_-}{\gamma_t}+1\right)}
{\Gamma\left(\frac{\o_++\o_-+d/ds}{\gamma_t}+1\right)}
\left(\tilde{F}_{\frac{\o_+}{\gamma_t},\frac{\o_-}{\gamma_t},t}(s)-\frac{d/ds}{\o_++\o_-}
\int_{-\infty}^{\infty}
du e^{-e^{\gamma_t(s-u)}}
g_{\gamma_t}(u)
\right)\notag\\
&=\frac{d/ds}{\Gamma(1+\gamma_t^{-1}d/ds)}
\int_{-\infty}^{\infty} du e^{-e^{\gamma_t(s-u)}}(\nu_{w,t}(u;X)
-\nu_{w,t}^{(\delta)}(u;X)).
\label{Fwlimit}
\end{align}
Here $\nu_{w,t}(u;X)$ and $\nu_{w,t}^{(\delta)}(u;X)$ are 
given by~\eqref{Result2}. 
In the last equality, we used the fact $\lim_{\substack{\o_-\rightarrow +\o_-\\ \o_+=w}}\tilde{F}_{\frac{\o_+}{\gamma_t},\frac{\o_-}{\gamma_t},t}(s;X)=0$, which follows from
the divergence of the term $\chi$ in~\eqref{tildechi} in the stationary limit
$\o_++\o_-\rightarrow 0$.

Finally we derive the representation in~\eqref{lrep}.
By using $B_{a,b,u}^{(1)}$ and $B_{a,b,u}^{(2)}$ defined in~\eqref{B(12)},
$L_{a,b}(u)$ is written as
\begin{align}
&L_{a,b}(u)
=\frac{1-a-b}{a+b}-\int_{-\infty}^{\infty}dx C_t(x)
B_{a,b,u}^{(1)}(x)B_{b,a,u}^{(1)}(x)\notag\\
&+\int_{-\infty}^{\infty}dxC_t(x)\left(B_{a,b,u}^{(1)}(x)B_{b,a,u}^{(2)}(x)
+B_{b,a,u}^{(1)}(x)B_{a,b,u}^{(2)}(x)\right)-\int_{-\infty}^{\infty}dx C_t(x) 
B_{a,b,u}^{(2)}(x)B_{b,a,u}^{(2)}(x).
\label{lex1}
\end{align}
Thus it is enough to show that in the limit $b\rightarrow -a$, 
the first two terms of this equation 
converge to $-2\gamma/\gamma_t+u-a^2-1$ in \eqref{lrep}
since the convergences
for the remaining terms are obvious.
Using the formula
\begin{align}
\int_{-\infty}^{\infty}C_t(x)e^{-yx} dx
=\frac{\pi}{\gamma_t}\cot\left(\frac{y\pi}{\gamma_t}\right)
\label{cotrelation}
\end{align}
for $0< \text{Re}(y)<\gamma_t$, we can further rewrite the term
$\int_{-\infty}^{\infty}dx C_t(x)
B_{a,b,u}^{(1)}(x)B_{b,a,u}^{(1)}(x)$ as
\begin{align}
-\int_{-\infty}^{\infty}dx C_t(x)
B_{a,b,u}^{(1)}(x)B_{b,a,u}^{(1)}(x)
=-\frac{e^{\frac{a^3+b^3}{3}-(a+b)u}\pi}
{\Gamma\left(\frac{a+b}{\gamma_t}+1\right)^2\gamma_t}
\cot\left(\frac{a+b}{\gamma_t}\pi\right)
\label{lex2}
\end{align}
for $0< \text{Re}(a+b)<\gamma_t$.
Considering the following properties,
\begin{align}
\Gamma(1+x)=1-\gamma x+O(x^2),~\cot(x)=\frac{1}{x}-\frac{x}{3}+O(x^2),
\end{align}
for $x\rightarrow 0$ with $\gamma$ being Euler's constant,
we easily find 
\begin{align}
\lim_{b\rightarrow -a}\frac{1-a-b}{a+b}-\int_{-\infty}^{\infty}dx C_t(x)
B_{a,b,u}^{(1)}(x)B_{b,a,u}^{(1)}(x)=-\frac{2\gamma}{\gamma_t}+u-a^2-1.
\label{limitb1b1}
\end{align}

We notice that although the function $L_{a,b}(u)$ should be valid for 
$0<a+b$ corresponding to the condition for the drifts $v_{\pm}$
discussed below~\eqref{aiggggex}, the more strict condition $0<a+b<\gamma_t$ 
is necessary for~\eqref{lex1} in order to 
apply the relation~\eqref{cotrelation} to~\eqref{lex2}
and other terms in~\eqref{lex2},
\begin{align}
&\int_{-\infty}^{\infty}dx C_t(x)
B_{a,b,u}^{(1)}(x)B_{b,a,u}^{(2)}(x)=
\frac{-\pi}{2\pi\gamma_t}\int_{\R+ic}dz 
\cot\left(\frac{a-iz}{\gamma_t}\pi\right)
\frac{e^{izu+\frac{i}{3}z^3}}{b+iz}
\frac{\Gamma\left(\frac{b+iz}{\gamma_t}+1\right)}
{\Gamma\left(\frac{a-iz}{\gamma_t}+1\right)}\notag\\
&\hspace{5cm}\times\frac{e^{-au+\frac{a^3}{3}}}{\Gamma\left(\frac{a+b}{\gamma_t}+1\right)}
\label{b1b2}
\\
&\int_{-\infty}^{\infty}dx C_t(x)
B_{a,b,u}^{(2)}(x)B_{b,a,u}^{(2)}(x)=
\frac{\pi}{(2\pi)^2\gamma_t}\int_{\R+ic_1}dz_1
\int_{\R+ic_2}dz_2  
\cot\left(\frac{-i(z_1+z_2)}{\gamma_t}\pi\right)\notag\\
&\hspace{5cm}\times\frac{e^{i(z_1+z_2)u+\frac{i}{3}(z_1^3+z_2^3)}}
{(a+iz_1)(b+iz_2)}
\frac{\Gamma\left(\frac{a+iz_1}{\gamma_t}+1\right)
\Gamma\left(\frac{b+iz_2}{\gamma_t}+1\right)}
{\Gamma\left(\frac{b-iz_1}{\gamma_t}+1\right)
\Gamma\left(\frac{a-iz_2}{\gamma_t}+1\right)}
,
\label{b2b2}
\end{align}
where $c$ ($c_1,~c_2)$ satisfies the conditions $b<c<b+\gamma_t$ 
and $0<a+c<\gamma_t$ ($a<c_1<a+\gamma_t$, $b<c_2<b+\gamma_t$ and 
$0<c_1+c_2<\gamma_t$). 
An advantage of this expression is that we can take the limit $b\rightarrow -a$
for $L_{a,b}(u)$ as in~\eqref{limitb1b1}. Moreover RHSs 
of~\eqref{lex2},~\eqref{b1b2} and~\eqref{b2b2} are analytic also for $a+b<0$
and thus so is $L_{a,b}(u)$~\eqref{lex1}. Noting 
the relations~\eqref{Result}--\eqref{Th3g} and considering RHS of~\eqref{Th3g} 
is represented by using~\eqref{stlimit1} and~\eqref{stlimit2}, we find 
${F}_{v_+,v_-,t}(s;X)$ is analytic also for the case $v_++v_-<0$. 

In the case $\gamma_t\le a+b$, on the other hand, 
another expression for $B_{a,b,u}(x)$
\begin{align}
B_{a,b,u}(x)=\int_{0}^{\infty}d\lambda e^{-a\lambda}\Ai_{\Gamma}^{\Gamma}
\left(x+u-\lambda,\frac{1}{\gamma_t},
1+\frac{b}{\gamma_t},1+\frac{a}{\gamma_t}\right),
\label{defBab}
\end{align}
is convenient. Using this, $L_{a,b}(u)$ is calculated as
\begin{align}
&L_{a,b}(u)=\frac{1-a-b}{a+b}-
\frac{\pi}{(2\pi)^2\gamma_t}\int_{\R+i\bar{c}_1}dz_1
\int_{\R+i\bar{c}_2}dz_2  
\cot\left(\frac{-i(z_1+z_2)}{\gamma_t}\right)\frac{e^{i(z_1+z_2)u+\frac{i}{3}(z_1^3+z_2^3)}}
{(a+iz_1)(b+iz_2)}\notag\\
&\hspace{4cm}\times\frac{\Gamma\left(\frac{a+iz_1}{\gamma_t}+1\right)
\Gamma\left(\frac{b+iz_2}{\gamma_t}+1\right)}
{\Gamma\left(\frac{b-iz_1}{\gamma_t}+1\right)
\Gamma\left(\frac{a-iz_2}{\gamma_t}+1\right)},
\end{align}
where $\bar{c}_1$ and $\bar{c}_2$ satisfy the conditions
$\bar{c}_1<a,~\bar{c}_2<b$ and $0<\bar{c}_1+\bar{c}_2<\gamma_t$. 

\section{Long-time limit}\label{lmt}
Here we give a proof of Theorem 3. Noticing the limiting behavior 
\begin{align}
\lim_{t\rightarrow\infty}e^{-e^{\gamma_tx}}={\bf 1}_{x\le0},
\end{align}
we find  the long-time limit of Theorem 2 becomes
\begin{align}
F_{w}(s;X)=\frac{d}{ds}\int_{s}^{\infty}du
\left(\nu_w(u;X)-\nu_w^{(\delta)}(u;X)\right)
=\nu_w^{(\delta)}(s;X)-\nu_w(s;X),
\end{align}
where $\nu_w(s;X)$ is defined by~\eqref{nuw} and $\nu_w^{(\delta)}(s;X)$
is given by
\begin{align}
\nu_w^{(\delta)}(s;X)=\det\left(1-\A^{(\delta)}\right)\L_{w-X,-w+X}^{(\delta)}(s)
+\det\left(1-\A^{(\delta)}-\D_{w-X,-w+X}^{(\delta)}\right).
\label{nuwd}
\end{align}
Here $\A^{(\delta)}$ and $\D_{a,-a}^{(\delta)}$ are defined in the same manner
as $\A$~\eqref{mcA} and $\D_{a,-a}$~\eqref{mcDw} respectively with $P_0(x)$ replaced 
by $P_0(x)-\delta(x)$. The function $\L_{a,-a}^{(\delta)}(s)$ is defined in terms
of $\L_{a,-a}(s)$~\eqref{mcLw} and $\B_{a,s}(x)$~\eqref{mcBw} as 
$\L_{a,-a}^{(\delta)}(s)=\L_{a,-a}(s)-\B_{a,s}(0)\B_{-a,s}(0)$.
Thus for the proof of Theorem 3, it is sufficient to show 
\begin{align}
\nu_w^{(\delta)}(s;X)-\nu_w(s;X)=\frac{d}{ds}\nu_w(s;X).
\label{nweq}
\end{align}

For this purpose, we focus on the expression in Proposition 1 and
show that~\eqref{Result} has two equivalent forms corresponding to
LHS and RHS of~\eqref{nweq}
in the simultaneous stationary and long-time limit 
(see~\eqref{fw1} and~\eqref{fw2} below). First, taking
the long-time limit in~\eqref{Result} with
$v_{\pm}=\o_{\pm}/\gamma_t$, 
we have
\begin{align}
&F_{\o_+,\o_-}(s;X):=\lim_{t\rightarrow\infty}
F_{{\o_+}/{\gamma_t},{\o_-}/{\gamma_t},t}(s;X)
=\left(1+\frac{1}{\o_++\o_-}\frac{d}{ds}\right)
\tilde{F}_{\o_+,\o_-}(s;X),\notag\\
&\tilde{F}_{\o_+,\o_-}(s;X):=\lim_{t\rightarrow\infty}
\tilde{F}_{{\o_+}/{\gamma_t},{\o_-}/{\gamma_t},t}(s;X)
=1-\int_{s}^{\infty}du g(u;X),
\label{Fexo}
\end{align}
where $g(u;X)$ is defined as the long time limit of 
$g_{\gamma_t}(u;X)$~\eqref{Th3g} with 
$v_{\pm}=\o_{\pm}/\gamma_t$ and has the expression as follows,
\begin{align}
g(s;X)=\det\left(1-P_s(B_{+-}-P_{\Ai}^{+-})P_s\right)
-\det\left(1-P_s B_{+-}P_s\right).
\label{gs}
\end{align}
Here $B_{+-}=\lim_{t\rightarrow\infty}B_{\gamma_t}^{\Gamma}$
and $P^{+-}_{\Ai}=\lim_{t\rightarrow\infty}P^{\Gamma}_{\Ai}$ are given by
\begin{align}
&B_{+-}(\xi_1,\xi_2)=\int_{0}^{\infty}dy
\Ai_{+-}(\xi_1+y,\o_+-X,\o_-+X)\Ai_{+-}(\xi_1+y,\o_-+X,\o_+-X),
\label{Bpm}
\\
&P^{+-}_{\Ai}(\xi_1,\xi_2)=\Ai_{+-}(\xi_1,\o_+-X,\o_-+X)
\Ai_{+-}(\xi_2,\o_-+X,\o_+-X),
\end{align}
where
\begin{align}
\Ai_{+-}(a,b,c)=\lim_{t\rightarrow\infty}\Ai_{\Gamma}^{\Gamma}\left(a,\frac{1}{\gamma_t},\frac{b}{\gamma_t},\frac{c}{\gamma_t}\right)=\int_{\Gamma_{ic}}dz
e^{iza+i\frac{z^3}{3}}\frac{b-iz}{c+iz}.
\end{align}
Next taking the stationary limit 
($\o_-\rightarrow -\o_+,~\o_+=w$) 
of $F_{\o_+,\o_-}(s;X)$~\eqref{Fexo}, we have
\begin{align}
\lim_{\substack{\o_-\rightarrow -\o_+\\ \o_+=w}}F_{\o_+,\o_-}(s;X)
=\lim_{\substack{\o_-\rightarrow -\o_+\\ \o_+=w}}\frac{1}{\o_++\o_-}g(s;X),
\label{Fopmslimit}
\end{align}
where we used the fact $\lim_{\substack{\o_-\rightarrow -\o_+\\ \o_+=w}}
\tilde{F}_{\o_+,\o_-}(x,X)=0$ as in~\eqref{Fwlimit}.

Now we notice that RHS of~\eqref{Fopmslimit} can be rewritten in two ways.
First, we use the results~\eqref{stlimit1} and~\eqref{stlimit2} 
in the previous section for the expression of $g(s;X)$~\eqref{gs} . By taking
the long-time limit for~\eqref{stlimit1} and~\eqref{stlimit2}, we see
\begin{align}
&\det\left(1-P_u(B_{+-}-P_{\Ai}^{+-})P_u\right)=
(\o_++\o_-)\nu_{\o_+,\o_-}^{(\delta)}(s;X),
\label{detnuod}\\
&\det\left(1-P_u B_{+-}P_u\right)
=(\o_++\o_-)\nu_{\o_+,\o_-}(s;X),
\label{detnuo}
\end{align}
where
\begin{align}
&\nu_{a,b}(s;X)=\det\left(1-\A\right)\L_{a+X,b-X}(s)
+\det\left(1-\A-
\D_{a+X,b-X}\right),
\label{nuo}
\end{align}
$\A(x,y)$, $\L_{a,b}(s)$, $\D_{a,b}(x,y)$ and $B_{a,s}(x)$ 
are defined by~\eqref{mcA}--\eqref{mcBw}.
The function $\nu_{\o_+,\o_-}^{(\delta)}(s;X)$ in~\eqref{detnuod} 
is defined in the same way as~\eqref{nuo} with $P_0$ replaced
by $P_0(x)-\delta(x)$. Note that
$\L_{a,b}(s)$ has the following expression 
\begin{align}
&\L_{a,b}(s)=\frac{1-e^{\frac{a^3+b^3}{3}-(a+b)s}}{a+b}-1
+\int_0^{\infty}dx\left(\B^{(1)}_{a,s}(x)\B^{(2)}_{b,s}(x)
+\B^{(1)}_{b,s}(x)\B^{(2)}_{a,s}(x)\right)\notag\\
&\hspace{5mm}-\int_0^{\infty}dx \B_{a,s}^{(2)}(x)\B_{b,s}^{(2)}(x),
\end{align}
which corresponds to~\eqref{lex1} and~\eqref{lex2}.
From~\eqref{mcA}--\eqref{mcBw},
one easily finds
\begin{align}
\lim_{\substack{\o_-\rightarrow -\o_+\\ \o_+=w}}\nu_{\o_+,\o_-}(s;X)
=\nu_w(s;X),
\label{stlnu}
~~
\lim_{\substack{\o_-\rightarrow -\o_+\\ \o_+=w}}
\nu_{\o_+,\o_-}^{(\delta)}(s;X)=\nu_w^{(\delta)}(s;X).
\end{align} 
Thus from~\eqref{gs},~\eqref{detnuo} and~\eqref{detnuod},
we find that RHS of~\eqref{Fopmslimit} is rewritten as
\begin{align}
\lim_{\substack{\o_-\rightarrow -\o_+\\ \o_+=w}}\frac{1}{\o_++\o_-}g(s;X)
=\nu_w^{(\delta)}(s;X)-\nu_w(s;X).
\label{fw1}
\end{align}

For the second expression for RHS of~\eqref{Fopmslimit}, 
we apply to~\eqref{gs} the relations
\begin{align}
&~~\det\left(1-P_s(B_{+-}-P_{\Ai}^{+-})P_s\right)\notag\\
&=\det\left(1-P_sB_{+-}P_s\right)
\left(1+\int_{-\infty}^\infty dx \left((1-P_sB_{+-}P_s)^{-1}P_s
P_{\Ai}^{+-}P_s\right)(x,x)\right),\\
&~\int_{-\infty}^\infty dx\left( (1-P_sB_{+-}P_s)^{-1}P_s
P_{\Ai}^{+-}P_s\right)(x,x)\notag\\
&=\frac{d}{ds}\text{tr}\log(1-P_sB_{+-}P_s)
=\frac{d}{ds}\log\det(1-P_sB_{+-}P_s).
\end{align}
As a result we find $g(s;X)$~\eqref{gs} is expressed as
\begin{align}
g(s;X)=\frac{d}{ds}\det(1-P_sB_{+-}P_s)=(\o_++\o_-)\frac{d}{ds}
\nu_{\o_+,\o_-}(s;X)
\label{gexo}
\end{align}
where in the second equality, we used~\eqref{detnuo}. Hence 
using~\eqref{stlnu}
we obtain the second expression for RHS of~\eqref{Fopmslimit},
\begin{align}
\lim_{\substack{\o_-\rightarrow -\o_+\\ \o_+=w}}\frac{1}{\o_++\o_-}g(s;X)
=\frac{d}{ds}\nu_w(s;X).
\label{fw2}
\end{align}
From~\eqref{fw1} and \eqref{fw2}, we find that~\eqref{nweq} holds.

At the end of this section, we briefly comment the equivalence between 
our representation in Theorem 3 and the one obtained in the studies of 
the PNG model~\cite{BR2000}. 
From~\eqref{Fexo} and the first equality in~\eqref{gexo}, 
we have the following expression
\begin{align}
F_{\o_+,\o_-}(s;X)=\lim_{t\rightarrow\infty}
F_{{\o_+}/{\gamma_t},{\o_-}/{\gamma_t},t}(s;X)
=\left(1+\frac{1}{\o_++\o_-}\frac{d}{ds}\right)
\det\left(1-P_sB_{+-}P_s\right).
\label{Fexo2}
\end{align}
Noticing that $B_{+-}(x)$~\eqref{Bpm} can be rewritten as
\begin{align}
B_{+-}(x,y)=\int_0^{\infty}d\lambda\Ai(x+\lambda)\Ai(y+\lambda)
+(\o_++\o_-)\B_{\o_++X,s}(x)\B_{\o_--X,s}(y),
\end{align}
we find that this expression is equivalent to the one in Theorem 5.1 
in~\cite{IS2004},
which is the limiting height distribution function in the PNG model 
with external sources. The relation between~\eqref{Fexo2} and 
the expression in~\cite{BR2000} was also discussed
in Proposition 5.2 in~\cite{IS2004}. This ensures that our result 
in Theorem 3 is equivalent to the representation in~\cite{BR2000}
in terms of the solution to the Painlev{\'e} II equation.
\section{Conclusion}\label{Co}
In this paper we have considered the one-dimensional KPZ 
equation~\eqref{KPZ} in the stationary situation, which 
corresponds to the growth from the two-sided Brownian
motion initial condition~\eqref{init_2bm}.
Combining the Bethe ansatz results of the one-dimensional attractive 
$\delta$-Bose gas with some techniques developed in this paper, such as 
the combinatorial identity and the shifting procedure, 
we have obtained the compact representation of the probability distribution 
of the height. Moreover the variance of this distribution provides 
the exact solution of the two-point correlation function.  This function 
is a most fundamental quantity in statistical physics
characterizing the space-time correlation and even if we focus only on the KPZ equation, it has been discussed by various approaches~\cite{FNS1977,CM2001,
KS2004,CCDW2011p}.  Our exact solution in this paper should help us 
understand the nontrivial space-time correlation 
for the KPZ equation and universality in a deeper way.

An important point for getting these exact results is to find out 
the Fredholm determinant structure for the generating 
function~\eqref{KPZfred1m}. However as pointed out in the introduction, 
the generating function is in fact a divergent sum. 
This can be confirmed by noticing from~\eqref{eigenvalue} that the ground state energy of the $N$-particle $\delta$-function attractive Bose gas is proportional to $-N^3$ and thus the $N$ replica partition function 
$\langle Z^N(x,t)\rangle$~\eqref{KPZdBose1} is proportional to $e^{N^3}$. 
We avoid this difficulty in the following way:
Using Lemma 6 (a), we rewrite $e^{N^3}$ in terms of $e^{N}$ and then
we take a partial sum with respect to $N$ in~\eqref{alternative1m}. 
(Note that some kind of analytic
continuation was used in this equation.)
Although the method involves some tricky procedures, we can say 
that our final results are fairly persuasive because of the following 
reasons.
\begin{enumerate}
\item The rigorous version of the replica analysis has been recently 
studied in a certain interacting particle process called the $q$-TASEP
of which an appropriate continuum limit provides the KPZ 
equation~\cite{BC2011p,BCS2012p}. 

\item As explained in Sec.~2, our result in Proposition 1 includes 
the exact solutions with narrow wedge~\cite{SS2010a,SS2010b,SS2010c,ACQ2010,
CLDR2010,D2010,D2010p2} and half BM~\cite{CQ2010p,IS2011} 
initial data as special cases. For both cases, the rigorous analyses by using
the ASEP~\cite{SS2010a,SS2010b,SS2010c,ACQ2010,CQ2010p} has been done and 
the equivalence between these results and the ones by 
the replica method~\cite{CLDR2010,D2010,D2010p2,IS2011} has
been confirmed.

\item In~LHS of~\eqref{lemma6(2)}, 
the new gamma function factor appears in addition to the factor $e^{m^3n^3/3}$, which is the origin of the divergence of the generating function 
in the case of narrow wedge initial data. However from the asymptotic 
property of the gamma function, we easily find the gamma function 
factor does not cause a singular behavior especially for 
large $n$. Thus the divergence of the generating function 
comes only from the factor $e^{m^3n^3/3}$,
which is the same situation as the narrow wedge initial data.
Moreover as explained in~\eqref{lemma6(2)},
Lemma 6(a), which plays a crucial role 
in the replica analysis, is a natural extension 
of the cases for the narrow wedge and half BM 
initial data.

\end{enumerate}
To understand the regularization procedure in the replica analysis more deeply
is an interesting future problem. We expect that
the findings in the replica approach to the KPZ equation should lead to the promotion of understanding the replica method in general disordered system.

\section*{Acknowledgments}
TS thanks  
A. Borodin, I. Corwin, P.L. Ferrari, S. Prolhac, J. Quastel and H. Spohn  for useful discussions.
Both authors would like to thank R.Y. Inoue 
for enjoyable conversations on related issues. 
The work of T.I. and T.S. is supported by 
KAKENHI (22740251) and KAKENHI (22740054) respectively.

\end{document}